\documentclass[a4paper,fleqn,usenatbib]{mnras}

\usepackage{newtxtext,newtxmath}

\usepackage[T1]{fontenc}
\usepackage{ae,aecompl}

\usepackage{graphicx}	
\usepackage{amsmath}	
\usepackage{amssymb}	
\usepackage{pdflscape}

\title[Soft X-ray and {[O\,III]} morphologies of AGN]{A comparison between the soft X-ray and {[O\,III]} morphologies of active galactic nuclei}

\author[C. G{\'o}mez-Guijarro et al.]{
Carlos G{\'o}mez-Guijarro,$^{1}$\thanks{E-mail: cgguijarro@dark-cosmology.dk}
Omaira Gonz{\'a}lez-Mart{\'i}n,$^{2}$
Cristina Ramos Almeida$^{3,4}$
\newauthor
Jos{\'e} Miguel Rodr{\'i}guez-Espinosa$^{3,4}$
and Jes{\'u}s Gallego$^{5}$
\\
$^{1}$Dark Cosmology Centre, Niels Bohr Institute, University of Copenhagen, Juliane Maries Vej 30, DK-2100 Copenhagen, Denmark\\
$^{2}$Instituto de Radioastronom{\'i}a y Astrof{\'i}sica (IRyA-UNAM), 3-72 (Xangari), 8701, Morelia, Mexico\\
$^{3}$Instituto de Astrof{\'i}sica de Canarias, C/V{\'i}a L{\'a}ctea, s/n, E-38205 La Laguna, Spain\\
$^{4}$Departamento de Astrof{\'i}sica, Universidad de La Laguna, E-38205 La Laguna, Spain\\
$^{5}$Departamento de Astrof{\'i}sica y CC. de la Atm{\'o}sfera, Facultad de CC. F{\'i}sicas, Universidad Complutense de Madrid\\ Av. Complutense s/n, E-28040 Madrid, Spain
}

\date{Accepted 2017 April 27. Received 2017 April 26; in original form 2016 October 16}

\pubyear{2017}

\begin{document}
\label{firstpage}
\pagerange{\pageref{firstpage}--\pageref{lastpage}}
\maketitle

\begin{abstract}
Several studies of nearby active galactic nuclei (AGN) have shown that the soft X-ray emission presents a size and morphology that resembles that of the narrow-line region (NLR) traced by [\ion{O}{iii}]. Since the NLR is mainly constituted by gas photoionised by the AGN, it seems logical to assume that this is also the primary source of the soft X-ray emission. However, these results are based on individual sources or small samples, particularly focused on type-2 Seyfert galaxies. Very little has been said concerning other types of AGN. The purpose of this work is to compare the circumnuclear morphologies of soft X-ray and [\ion{O}{iii}] images to test whether they match in different optical classes of AGN. Our sample is composed of 27 AGN: nine type-1 Seyferts, 10 type-2 Seyferts, and eight low ionisation nuclear emission-line regions (LINERs). We find a good match in 100\% of the type-2 Seyferts in our sample. This correspondence is less frequent in type-1 Seyferts (22\%) and it is not seen in LINERs. The good resemblance in type-2 Seyferts constitutes an evidence for a common physical origin. We argue that the lack of correspondence in type-1 Seyferts might be due to the line of sight perpendicular to the accretion disk. Based on the morphologies of the eight LINERs in our sample, we discard a common origin for the soft X-ray and [\ion{O}{iii}] emissions in these objects. Regarding the X-ray properties, both high column density and hard X-ray luminosity are associated with matched morphologies.
\end{abstract}




\section{Introduction}
\label{sec:intro}

The unified model (UM) of active galactic nuclei (AGN) explains the observational properties of a particular AGN simply by its luminosity and orientation \citep{1993ARA&A..31..473A,1995PASP..107..803U}. Among AGN, Seyfert galaxies \citep{1943ApJ....97...28S} can be classified as type-1 or type-2 depending on the presence or not of permitted broad lines in their optical spectra. The UM proposes that both types of Seyferts are essentially the same objects viewed at different angles. Type-1 Seyfert galaxies are observed in an viewing angle in which the line of sight (LOS) is able to reach the region where these broad emission lines are produced, the broad line region (BLR). This BLR is probably partially absorbed by photoionised outflowing material launched by the accretion disk \citep{1994PASJ...46L..59F,1995MNRAS.273.1167R}. Type-2 Seyferts are observed at an angle nearly perpendicular to the type-1 orientation. From this view, the nuclear continuum is highly obscured by an optically thick dusty structure (simplified as a torus) blocking the BLR from our LOS.

A strong observational evidence for unification between type-1 and type-2 Seyferts was the discovery of broad optical lines in the polarised spectrum of the archetypal type-2 Seyfert M77 \citep[NGC\,1068,][]{1985ApJ...297..621A}, an evidence further supported by the last compilation of spectropolarimetry data of type-2 Seyfert galaxies \citep{2016MNRAS.461.1387R}. In addition, X-ray wavelengths are also affected by the obscuring material along the LOS. They produce the best estimate of the amount of material measuring the hydrogen column density ($N_{\rm{H}}$). X-ray observations of AGN provided additional evidence in favour of the UM, with type-2 AGN showing larger column density in our LOS, as expected under the UM \citep{1998A&A...338..781M,1999ApJ...522..157R,2006A&A...455..173P,2006A&A...446..459C}.

The hard X-ray energies (i.e., 2.0-10.0\,keV) are thought to originate in the innermost regions of the accretion flow which feeds the AGN. Thus, they are a good tracer of the AGN power \citep{2003MNRAS.344L..59M}. For these energies, the spectrum is dominated by an intrinsic continuum, and by Compton reflection and a strong iron fluorescence line coming from reflected light from the AGN in the inner parts of the dusty structure and/or in the accretion disk \citep{1998A&A...338..781M,2005A&A...444..119G}. This Compton reflection is particularly prominent for Compton-thick (CT) objects (i.e., $N_{\rm{H}} > 1.5 \times 10^{24}$\,cm$^{-2}$) due to the observational faintness of the intrinsic continuum \citep{1999ApJS..121..473B}.

The origin of the soft X-ray energies (i.e., 0.2-2.0\,keV) have been more controversial in both unobscured and obscured sources. In the former is dominated by an excess in the primary nuclear continuum of debated origin. In obscured sources, where the primary emission is attenuated, its nature was partially revealed through the high resolution spectra obtained by \emph{XMM-Newton} and \emph{Chandra} spectrometers for some type-2 Seyferts \citep[e.g., Mrk\,3, Circinus, and NGC\,1068, by][]{2000ApJ...543L.115S,2001ApJ...546L..13S,2002ApJ...575..732K,2002A&A...396..761B}. In these studies the spectrum is dominated by emission lines, mainly from He- and H-like K transitions of light metals and L transitions of Fe, that are blended in lower resolution spectra. This is consistent with photoionised material from the AGN.

Indeed, the AGN is able to photoionised the material in its host galaxy up to kpc-scales \citep{1993ARA&A..31..473A}. The more extended regions are the so-called narrow-line region (NLR) and extended narrow-line region (ENLR), observed in many nearby Seyfert galaxies \citep{2003ApJS..148..327S,2003AJ....126.2185V,2004AJ....127..606W,2006ApJ...645..148R,2006A&A...448..499B,2009ApJ...694..718W,2010ApJ...723.1748G} and at higher redshifts in radio galaxies \citep{2012A&A...545A.143B}. The study of the NLR/ENLR is key to impose restrictions on the interaction between the AGN and its host because it is the outer layer powered by the AGN. AGN drive jets and winds that have the potential to directly influence the cooling of the hot gas in host galaxy/cluster haloes \citep{2007ARA&A..45..117M}, and also to expel cool gas from the circumnuclear regions of the galaxies. Thus, studying the NLR can give hints on the AGN-host interaction and feedback as a key element in the understanding of both galaxy and AGN evolution \citep{1997ApJS..113...23T,1998A&A...331L...1S,2003MNRAS.346.1055K,2006ApJ...652..864H,2007MNRAS.382.1415S,2009ApJ...692L..19S,2012ARA&A..50..455F}.

Observationally, this region has been classically traced by the [\ion{O}{iii}]$\lambda$5007 emission line \citep{2003ApJS..148..327S,2003ApJ...597..768S}, with heterogeneous extensions (hundreds of parsecs to kpc-scales), column densities ($N_{\rm{H}} \sim 10^{19}$--$10^{21}$\,cm$^{-2}$), and morphologies showing halo-like, cone-like, spiral or amorphous features. Conical or bi-conical morphologies suggest the origin of the NLR being outflowing material coming from the very center of the galaxy, probably launched by the accretion disk and, perhaps, collimated by the torus \citep{2000ApJ...530L..65N}. \citet{2003ApJ...597..768S} found that type-1 Seyferts show a high percentage of halo-like NLRs with a high concentration towards the nucleus, while those of type-2 are more elongated and less concentrated. This suggests that the outflowing material main direction might be parallel to our LOS in type-1 Seyferts while it might be perpendicular in type-2 Seyferts, in agreement with predictions from the UM.

The soft X-ray energy band gives a new opportunity to infer the properties of the NLR/ENLR \citep{1999ApJ...516..750C,2003ARA&A..41..117C}. Diagnostics of the emission lines in the spectra showed that they are more easily explained if produced in a gas photoionised by the AGN, rather than in hot gas in collisional equilibrium in type-2 Seyferts \citep{2007MNRAS.374.1290G}. The high spatial resolution of \emph{Chandra} revealed that the soft X-ray emission shows a size and morphology that closely resembles that of the NLR traced by the [\ion{O}{iii}] emission for several nearby type-2 Seyferts, as several studies showed from individual objects: Mrk\,3 \citep{2000ApJ...543L.115S}, NGC\,1068 \citep{2001ApJ...556....6Y}, NGC\,2110 \citep{2006ApJ...653.1121E}, Mrk\,573 \citep{2010MNRAS.405..553B}, and from a small sample of type-2 Seyferts \citep[8 type-2 Seyferts, ][]{2006A&A...448..499B}. The good match between both soft X-ray and [\ion{O}{iii}] morphologies was also reported in the type-1 Seyfert NGC\,4151 \citep{2000ApJ...545L..81O,2001ApJ...563..124Y}. Since the NLR is mainly constituted by gas photoionised by the AGN, it seems then logical to assume that this is also the primarily source of the soft X-ray emission. However, other studies suggested that a continuum-subtracted [\ion{O}{iii}] does not match the morphology of the soft X-rays, and thus, the physical mechanism driven its emission should be different \citep[Mrk\,1066,][]{2014MNRAS.445.1130R}. Soft X-rays can also be produced through mechanical heating, as in shocks driven by supernova explosions in nuclear star-forming regions, or by outflowing material \citep{2000ApJ...544..763K,2002ApJ...572..753D}. Star formation in the disks and galactic winds in hot halos of spiral galaxies can be traced by the diffuse X-ray emission as well \citep{2004ApJ...610..213T,2004ApJS..151..193S,2006ApJ...647.1030S,2007MNRAS.376.1611W,2009MNRAS.394.1741O,2010ApJS..188...46K,2012MNRAS.426.1870M}. It is quite plausible that both photoinisation and mechanical heating effects are important \citep[Mrk\,3,][]{2000ApJ...543L.115S}. In other cases, the diffuse X-ray emission can be explained only through a thermal origin \citep[NGC\,1365,][]{2009ApJ...694..718W}.

The UM has been vastly tested for Seyfert galaxies with a good general agreement adding some considerations, as in studies of the Seyfert classification depending on the torus structure \citep{2011ApJ...731...92R}. However, some other AGN types are not easily explained under the UM \citep[][for a review]{2012AdAst2012E..17B}. One clear example are the \emph{low ionisation nuclear emission-line regions} \citep[LINERs, ][]{1980A&A....87..152H}, originally defined as a subclass of \emph{low-luminosity} AGN (LLAGN). They show optical spectra dominated by emission lines of moderate intensities arising from gas in lower ionisation states than classical AGN. In the X-ray regime, several studies proved the AGN nature in a large fraction of LINERs \citep{2001ApJ...549L..51H,2002ApJ...565..108E,2005ApJ...620..113D,2006A&A...460...45G,2009A&A...506.1107G}. Previous studies of LINERs reached different conclusions about the ionisation mechanism responsible for the LINER emission. Possibilities included: 1) shock heating \citep{1995ApJ...455..468D}, 2) Wolf-Rayet or OB stars in compact near-nuclear star clusters \citep{1985MNRAS.213..841T,1992ApJ...397L..79F}, and 3) low-luminosity AGN \citep{1997ApJS..112..391H,2001ApJ...554..240E}. It was suggested that LINERs are low-luminosity, low-accretion rate extension of Seyferts \citep{2008ARA&A..46..475H}. Only one previous attempt was done to study the NLR/ENLR in LINERs \citep{2011A&A...527A..23M}. They found extended emission in the soft X-rays with equivalent sizes ranging from a few tens to about 500\,pc that resembled the optical morphologies. However, the tracer used there was H$\alpha$, which is not the best suited to trace the NLR/ENLR region due to possible contamination of star formation.

The main purpose of this paper is to study the correspondence between the soft X-ray and [\ion{O}{iii}] emission-line morphologies in a large sample of nearby AGN with different optical classes. The sample includes nine type-1 Seyferts, 1 type-2 Seyferts, and eight LINERs. We used both [\ion{O}{iii}] emission-line images taken with the \emph{Hubble} Space Telescope (\emph{HST}) and X-ray images from the \emph{Chandra} satellite. The high angular resolution provided by \emph{HST} and \emph{Chandra} is optimal to trace the morphology of the NLR.

The paper is structured as follows. We introduce the sample in Section~\ref{sec:sample}. In Section~\ref{sec:data} we describe the \emph{HST} and \emph{Chandra} data employed along with its treatment, and the methodology in Section~\ref{sec:method}. We present our results in Section~\ref{sec:results} and discuss them in Section~\ref{sec:discussion}. Summary and conclusions are placed in Section~\ref{sec:summary}. Along this paper we adopted a concordance cosmology $[\Omega_\Lambda,\Omega_M,h]=[0.7,0.3,0.7]$.

\section{Sample}
\label{sec:sample}

We employed several catalogues of AGN to select objects suitable to our analysis: the Palomar Survey \citep{1997ApJS..112..315H}, the \citet{2010A&A...518A..10V} catalogue, and the extended \emph{IRAS} 12 micron galaxy sample \citep[12MGS,][]{1993ApJS...89....1R}. From them, we gathered a sample of 27 sources with both \emph{Chandra} and [\ion{O}{iii}] data available in the archives.

The sample includes several AGN optical types: nine Sy1-1.9, 10 Sy2 galaxies, and eight LINERs. Note that from now on we refer to Sy1-1.9 as type-1 Seyferts and Sy2 as type-2 Seyferts, denoting the existence or not of broad components in the optical permitted lines. This is the largest sample of AGN ever compiled to compare \emph{Chandra} and optical morphologies. We broaden the previous studies focused on type-2 Seyferts to a wide variety of optical classes of AGN. It is worth noticing that our sample is subjected to the individual reasons for which they were observed with \emph{Chandra} and \emph{HST}. Thus, it is not complete and the biases are out of the control of the present study. \citet{2006A&A...448..499B} reported a sample of eight type-2 Seyferts to also compare the soft X-ray and [\ion{O}{iii}] morphologies. Among the 10 type-2 Seyferts presented here, six are in common with their sample (NGC\,1386, NGC\,3393, NGC\,4388, NGC\,4507, NGC\,5347, and NGC\,5643).

In the case of the LINERs it is important to indicate those reported as having proper AGN activity, showing features characteristic of AGN such as: evidence of a BLR; radio jets or unresolved radio core; or ultraviolet variability. This is the case for the sources listed here: M31 \citep{2005ApJ...631..280B}, M81 \citep{2000ApJ...532..895B}, M90 \citep{1983ApJ...269..466K}, M105 \citep{2009ApJ...704.1570G}, NGC\,404 \citep{2007MNRAS.377.1696M}, NGC\,1052 \citep{2009A&A...506.1107G}, NGC\,4278 \citep{1997ApJS..112..391H}, and NGC\,7465 \citep{2009ApJ...694.1379R}.

In Table~\ref{tab:sample}, we show the optical classification of the galaxies, the redshifts, the intrinsic hard X-ray luminosity and hydrogen column density collected from the literature, and the [\ion{O}{iii}] luminosities calculated in the nuclear region and also considering the extended emission. The methodology for obtaining the [\ion{O}{iii}] luminosities is explained in Section~\ref{sec:oiii_lum}. In addition, we include both X-ray and optical morphological classification, as explained in Sections~\ref{sec:xray_morph} and \ref{sec:oiii_morph}. Besides, to provide a quick look of the results, we show a checkmark when the source presents a good match between the soft X-ray and the [\ion{O}{iii}] morphologies.

\begin{landscape}
\begin{table}
\caption{AGN sample. `Sy' denotes Seyfert. References for the optical classification in Col.~2 are: (1) \citet{1976MNRAS.176P..61O}; (2) \citet{1997ApJS..112..315H}; (3) \citet{2006A&A...455..773V}; (4) \citet{1995PhDT........25H}; (5) \citet{1998ApJ...493..650M}; (6) \citet{1990MNRAS.245..749S}; (7) \citet{1993ApJ...414..552O}; (8) \citet{1995ApJS...99...67N}; (9) \citet{1997ApJ...487..122F}; (10) \citet{1986A&A...156...51D}; (11) \citet{2010A&A...518A..10V}; (12) \citet{1983ApJ...269..466K}; (13) \citet{2009ApJ...694.1379R}. References for the intrinsic hard X-ray luminosity and the hydrogen column density ($10^{22}$\,cm$^{-2}$) included in Cols.~4 and 5 are: (14) \citet{2011MNRAS.413.1206B}; (15) \citet{2006A&A...455..173P}; (16) \citet{2012ApJ...748..130M}; (17) \citet{2005MNRAS.358.1423O}; (18) \citet{1999ApJS..121..473B}; (19) \citet{2015ApJ...815....1U}; (20) \citet{2015ApJ...805...41B}; (21) \citet{2016MNRAS.455L..26X}; (22) \citet{2015A&A...579A..90H}; (23) \citet{2007ApJ...664..277Y}; (24) \citet{2004MNRAS.355..297G}; (25) \citet{2009A&A...506.1107G}; (26) \citet{2004A&A...414..825S}; and (27) \citet{2001MNRAS.328..461O}. Optical [\ion{O}{iii}] luminosities calculated in the nuclear region are denoted as `nuc' in Col.~6, while considering the nuclear and the extended regions of the source are indicated as `tot' in Col.~7. Luminosities are presented as the decimal logarithm of the values in CGS units. In Col.~8, X-ray (soft/hard) morphologies are shown: `Cone' denotes cone-shaped, `Sph.' stands for spheroidal, `Other' suggests diffuse morphologies that can not be classified into cone-shaped or spheroidal, `Diff.' indicates diffuse, and `Point' denotes point-like sources. The optical morphologies are gathered in Col.~9. The tick in Col.~10 refers to the presence of a good match between the soft X-ray and the [\ion{O}{iii}] morphologies.}
\label{tab:sample}
\begin{tabular}{l c c c c c c c c c}
\hline
Name & Optical type & $z$ & $\log {(L_{\rm{HX}})}$ & $N_{\rm{H}}$ & $\log {(L_{\rm{[\ion{O}{iii}],nuc}})}$ & $\log {(L_{\rm{[\ion{O}{iii}],tot}})}$ & X-ray morph. & Optical morph. & Match \\
\hline
IC\,450 (Mrk\,6) & Sy1.5 (1) & 0.0188 & 43.1 (14) & 2.91 (14) & 39.7 & 39.9 & Sph./Point & Sph. &  \\
M106 (NGC\,4258) & Sy1.9 (2) & 0.0017 & 40.9 (15) & 7 (15) & 36.9 & 37.6 & Sph./Point & Sph. &  \\
NGC\,931 & Sy1 (3) & 0.0116 & 43.4 (16) & 0.363 (16) & 38.5 & 38.7 & Sph./Point & Sph. &  \\
NGC\,1365 & Sy1.8 (3) & 0.0042 & 42.5 (14) & 17.5 (14) & 37.3 & 39.0 & Other/Point & Other & $\checkmark$ \\
NGC\,1569 & Sy1.5 (4) & 0.0007 & 38.5 (17) & 0.286 (17) &   & 38.7 & Sph./Point & Other &  \\
NGC\,4051 & Sy1.2 (2) & 0.0033 & 41.5 (14) & 18.5 (14) & 38.1 & 38.2 & Sph./Point & Cone &  \\
NGC\,4151 & Sy1.5 (3) & 0.0030 & 42.1 (14) & 5.96 (14) & 39.8 & 40.5 & Cone/Point & Cone & $\checkmark$ \\
NGC\,4395 & Sy1.8 (3) & 0.0010 & 39.8 (15) &   & 36.9 & 37.1 & Sph./Point & Cone &  \\
NGC\,5273 & Sy1.5 (2) & 0.0037 & 41.4 (15) &   & 37.2 & 37.7 & Cone/Point & Cone &  \\
\cline{1-10}
Circinus & Sy2 (5) & 0.0010 & 41.9 (18) & $>150$ (18) & 35.8 & 37.1 & Other/Point & Cone & $\checkmark$ \\
IC\,5063 & Sy2 (6) & 0.0088 & 42.8 (19) & 25 (19) & 38.3 & 39.6 & Cone/Point & Cone & $\checkmark$ \\
M51a (NGC\,5194) & Sy2 (3) & 0.0018 & 40.6 (20) & 700 (20) & 34.8 & 37.5 & Cone/Diff. & Cone & $\checkmark$ \\
M77 (NGC\,1068) & Sy2 (7) & 0.0030 & 42.4 (14) &   & 39.1 & 39.9 & Other/Diff. & Other & $\checkmark$ \\
NGC\,1386 & Sy2 (8) & 0.0038 & 41.8 (21) & 561 (21) & 38.0 & 38.7 & Cone/Diff. & Cone & $\checkmark$ \\
NGC\,3393 & Sy2 (9) & 0.0125 & 41.3 (22) & 24.3 (22) & 38.2 & 39.8 & Cone/Point & Cone & $\checkmark$ \\
NGC\,4388 & Sy2 (2) & 0.0048 & 42.9 (14) & 39.7 (14) & 37.4 & 38.7 & Cone/Point & Cone & $\checkmark$ \\
NGC\,4507 & Sy2 (10) & 0.0118 & 42.8 (22) & 41.8 (22) & 39.5 & 39.8 & Cone/Point & Cone & $\checkmark$ \\
NGC\,5347 & Sy2 (2) & 0.0064 & 42.5 (23) & $>150$ (23) & 37.5 & 38.3 & Cone/Point & Cone & $\checkmark$ \\
NGC\,5643 & Sy2 (2) & 0.0039 & 41.4 (24) & 77 (24) & 38.3 & 39.0 & Cone/Diff. & Cone & $\checkmark$ \\
\cline{1-10}
M31 (NGC\,224) & LINER (11) & 0.0002 & 38.8 (14) &   &   & 35.2 & Cone/Point & Sph. &  \\
M81 (NGC\,3031) & LINER (4) & 0.0009 & 40.2 (15) &   & 39.8 & 41.6 & Sph./Point & Sph. &  \\
M90 (NGC\,4569) & LINER (12) & 0.0029 & 37.7 (14) &   & 37.8 & 38.0 & Cone/Point & Other &  \\
M105 (NGC\,3379) & LINER (2) & 0.0025 & 38.1 (25) &   &   & 37.5 & Sph./Point & Sph. &  \\
NGC\,404 & LINER (3) & 0.0009 & 37.3 (26) & 0.053 (26) & 36.0 & 36.6 & Cone/Point & Sph. &  \\
NGC\,1052 & LINER (2) & 0.0046 & 41.5 (14) & 30.6 (14) & 40.7 & 42.2 & Cone/Point & Sph. &  \\
NGC\,4278 & LINER (2) & 0.0038 & 39.2 (25) & 2.65 (25) &   & 37.3 & Sph./Point & Sph. &  \\
NGC\,7465 & LINER (13) & 0.0064 & 41.4 (27) &   & 37.6 & 39.2 & Sph./Point & Sph. &  \\
\hline
\end{tabular}
\end{table}
\end{landscape}

The \emph{HST} instruments and filters queried to look for suitable objects for our analysis are specified in Table~\ref{tab:filters}. The spatial resolution of the \emph{HST} Wide Field Camera (WFC3, WFPC2, and WFPC) is the best for the purpose of this work. Several narrow-band (NB) filters were chosen to trace the [\ion{O}{iii}]$\lambda$5007, while the contiguous broad-band (BB) F547M filter was selected to subtract the local continuum below the [\ion{O}{iii}] line.

\begin{table}
\caption{\emph{HST} filter pairs employed in our analysis along with the redshift range that we can cover with each of them.}
\label{tab:filters}
\begin{tabular}{c c c c}
\hline
Instrument & NB & BB & Redshift range \\
\hline
WFC3 & F502N & F547M & $z<0.0070$ \\
WFC3 & FQ508N & F547M & $0.0037<z<0.0167$ \\
WFPC2 & F502N & F547M & $z<0.0037$ \\
WFPC2 & FR533N & F547M & $z<0.0195$ \\
WFPC & F502N & F547M & $z<0.0051$ \\
\hline
\end{tabular}
\end{table}

We selected objects with the [\ion{O}{iii}] emission line lying within the FWHM of the NB. This ensures enough strength in the [\ion{O}{iii}] emission for our analysis, avoiding biases in the results due to spurious detections in the edges of the filters. We calculated the percentage of the [\ion{O}{iii}] emission-line flux within each filter using the filter profiles and assuming that the [\ion{O}{iii}] emission line has a width of 200\,km s$^{-1}$. The [\ion{O}{iii}] images contain at least 90\% of the emission-line flux for all the sources, except for NGC\,4388 which still corresponds $\sim$85\% of the emission-line flux. Note that we did not consider the sources with [\ion{O}{iii}] emission line lying on the BB filter, otherwise we would subtract [\ion{O}{iii}] features that are precisely our aim. All these constraints establish a redshift range for each filter that allowed us to look for the objects in the catalogues through a redshift criterion. This redshift range for each \emph{HST} camera is shown in Table~\ref{tab:filters} (Col.~4). In the case of WFPC2 FR533N we just include the upper limit introduced by the BB filter because it is a tunable filter designed to trace the [\ion{O}{iii}] emission of the selected objects.

Once we had a raw sample of objects accomplishing the redshift criterion, we searched for the optical and X-ray data in the archives. For the optical data, we used the \emph{Hubble} Legacy Archive\footnote{http://hla.stsci.edu/} (HLA) DR8.1 version when available and The Mikulski Archive for Space Telescopes\footnote{https://archive.stsci.edu/} (MAST) otherwise. After that, we searched for the raw sample in the NASA archive for high energy mission HEASARC\footnote{http://heasarc.gsfc.nasa.gov/}. \emph{HST} and \emph{Chandra} images were available for 30 sources. We discarded three objects with major problems in the images that prevented us from performing the soft X-ray/[\ion{O}{iii}] comparison: one object that did not show any feature in the \textit{HST} images due to lack of sensitivity in the observations (NGC\,253), and two more objects that were completely dominated by pile-up in the \textit{Chandra} images (Centaurus A and M87). The final sample is comprised of 27 AGN.

We only used \emph{Chandra} X-ray images because its superb spatial resolution is the only one which allowed us to compare with optical morphologies probed by the \emph{HST}. The spatial scales range between 0.23\,pc/pix and 24\,pc/pix in the case of \emph{Chandra} (sub-pixel binning scale), with an average and standard deviation of $(5.5\pm5.5)$\,pc/pix (see Col.~11 in Table~\ref{tab:log} for the values of each source). For \emph{HST} the spatial scale range between 0.15\,pc/pix and 25\,pc/pix (drizzled pixel scale), with an average and standard deviation of $(6.0\pm6.9)$\,pc/pix (see Col.~6 of Table~\ref{tab:log} for each source).

\section{Data processing}
\label{sec:data}

\subsection{\emph{HST} data}
\label{sec:hst}

We used NB and BB images of several filter pairs, as described in Section~\ref{sec:sample}. When several observations were accesible through the archives for the same source, we took the data with the greatest exposure time, rejecting images with saturated pixels when possible. We refrained from combining several images to increase the signal to noise (S/N) ratio. In some cases one of the images was very noisy compared to the one used in our analysis. Thus, the combination was not able to produce a better result. In other cases, the two images were observed with different instruments with different pixel scales. Also different filter responses would affect the combination. The number of images where we could do a combination of images safely are only a one or two cases. We decided not to do it to avoid including errors in our comparison with the soft X-rays.

The majority of the data was retrieved from the HLA archive, except for NGC\,1052, and NGC\,3031 from the MAST archive. The HLA archive provides enhanced data products. Among them, the mosaic, combined image of different exposures, and flux calibration for each CCD. MAST archive does not provide the WFPC mosaic. In these cases, we only used the camera where it was located the area of interest. In Table~\ref{tab:log}, we include the detail of the observations extracted from the archives.

We removed the cosmic rays in affected images using the L.A.Cosmic\footnote{http://www.astro.yale.edu/dokkum/lacosmic/} algorithm. This algorithm uses a variation of the Laplacian edge detection method to remove cosmic rays \citep{2001PASP..113.1420V}. We then applied the flux calibration provided by the PHOTFLAM keyword included in the image headers. The next step consisted on the alignment of both NB and BB images if necessary. That was the case in nine objects (namely IC\,450, IC\,5063, NGC\,404, NGC\,1365, NGC\,1386, NGC\,3393, NGC\,4388, NGC\,4395, and NGC\,5347). We used stars in the field when available and compact features within the source otherwise.

Once we had NB and BB images properly treated and aligned, we performed the continuum subtraction. As a first approach, we subtracted the BB image from the NB one. However, visual inspection showed that this process failed to properly subtract the host galaxy continuum. It turned out in an over-subtraction in some cases (showing deep negative cavities in the images) and under-subtraction (with strong contribution of the host galaxy) in other cases. The over-subtraction was particularly important, as negative pixel values have no physical meaning. We wanted to make a more accurate subtraction of the continuum host by ensuring that the final subtracted image was scaled to zero around and far from the target source. Thus, the final image was neither under- nor over-subtracted. Technically the process was: (1) we selected several regions, $r_{i}$, common in both the NB and BB images; (2) we calculated the average flux, $\rm{<F_{NB}(r_{i})>}$ and $\rm{<F_{BB}(r_{i})>}$, respectively; (3) we computed  $\rm{F_{subtract}=<F_{NB}(r_{i})> - scl_{j} \times  <F_{BB}(r_{i})>}$, where $\rm{scl_{j}}$ is an array of plausible scaling values between the two; (4) for each $\rm{scl_{j}}$ value we computed the average $\rm{F_{subtract}}$ for all the regions $r_{i}$ ($\rm{<F_{subtract}>}$); (5) we plotted $\rm{<F_{subtract}>}$ versus $\rm{scl_{j}}$; (6) the final scaling value $\rm{scale}$ was where $\rm{<F_{subtract}>}$ was equal to zero. Note that the regions were selected far from the NLR but close enough to account for the continuum of the galaxy. We then guaranteed that the image background was set to zero level. The final continuum-subtracted [\ion{O}{iii}] image is:

\begin{equation}
 \mathrm{[\ion{O}{iii}]}=\mathrm{NB}-\mathrm{scale}\times\mathrm{BB} , \label{eq:oiii}
\end{equation}
\noindent where $\mathrm{scale}\sim1$ for all the objects. This was what we expected from a small correction. We chose this formula with the $\rm{scale}$ being a multiplier factor because the major source of error was expected to be the error in the flux calibration. Note that the resulting flux units are erg s$^{-1}$ cm$^{-2}$ $\AA^{-1}$.

\subsection{\emph{Chandra} data}
\label{sec:chandra}

We used the level 2 event data from the \emph{Chandra}/ACIS instrument retrieved from the HEASARC archive. When several observations were accesible for the same object, we chose the data with a largest exposure time, avoiding at the same time images with pile-up when possible. Pile-up occurs when the source is very bright and the X-ray CCD detector is not able to distinguish between photon arrivals at the same position between two read-outs, affecting the final counts value. We found pile-up in M77, NGC\,3031, and NGC\,4151, but just in a very compact region in the nucleus that does not affect the extended morphologies. To ensure low distortions, all our objects were observed with an offset between the optical axis and the source lower than 3\arcmin. We include the detail of the observations collected from the archive in Table~\ref{tab:log}.

The data reduction followed the process previously applied in \citet{2010ApJ...723.1748G} \citep[see also][]{2009A&A...506.1107G}. We employed the CIAO 4.7\footnote{http://cxc.cfa.harvard.edu/ciao/} data analysis system and the \emph{Chandra} Calibration Database\footnote{http://cxc.harvard.edu/caldb/} (caldb 4.6.5). The exposure time was processed to exclude background flares using the {\sc lc\_clean.sl} task\footnote{http://cxc.harvard.edu/ciao/download/scripts/} in source-free sky regions of the same observation. \emph{Chandra} data include information about the photon energies and positions that is used to obtain energy-filtered images and to carry out sub-pixel resolution spatial analysis. Although the default pixel size of the \emph{Chandra}/ACIS detector is 0.492\arcsec, smaller spatial scales are accessible as the image moves across the detector pixels during the telescope dither, therefore sampling pixel scales smaller than the default pixel of \emph{Chandra}/ACIS detector. This allows sub-pixel binning of the images. Similar techniques were applied for the analysis of \emph{Chandra} observations of, for example, the SN1987A remnant \citep{2000ApJ...543L.149B}. For each source in our sample, we used a value of 0.06\arcsec/pix, corresponding to 1/8 of the native pixel scale \citep[e.g.,][]{2011ApJ...729...75W,2012ApJ...756...39P}, better matching the pixel scale of the \textit{HST} images. In addition, we applied smoothing techniques to detect the low-contrast diffuse emission. We employed the CIAO tool {\sc csmooth}, based on the algorithm developed by \citet{2006MNRAS.368...65E}. {\sc csmooth} is an adaptive smoothing tool for images containing multi-scale complex structures and it preserves the spatial signatures and the associated counts, as well as significance estimates. A minimum significance S/N level of 2, and a scale maximum of 3 pixels was chosen for all the sources. For our analysis, we generated two images in the energy range of the soft X-rays (0.2-2.0\,keV) and the hard X-rays (2.0-10.0\,keV).

\begin{landscape}
\begin{table}
\caption{Log of the \emph{HST} and \emph{Chandra} observations employed in our analysis. Note that in all cases the BB filter is the F547M of the \emph{HST} instrument for each case. `Prop.' denotes proposal, `Obs.' observation, `Inst.' instrument, 1\,pix indicates the pixel scale in pc (sub-pixel binning scale), and `Exp.' is the exposure time in s for \emph{HST} data and ks for \emph{Chandra}. We show the proposal ID and the exposure time in the optical data for both the NB (Prop. ID NB, Exp. NB) and the BB (Prop. ID BB, Exp. BB).}
\label{tab:log}
\begin{tabular}{l | c c c c c c c | c c c c}
\hline
Name & \multicolumn{7}{c}{\emph{HST}} & \multicolumn{4}{c}{\emph{Chandra}} \\
 & Prop. ID NB & Prop. ID BB & Inst. & NB & 1\,pix & Exp. NB & Exp. BB & Obs. ID & Inst. & 1\,pix & Exp. \\
 & & & & & (pc) & (s) & (s) & & & (pc) & (ks)\\
\hline
Circinus & 7273 & 7273 & WFPC2/PC & F502N & 0.94 & 1800 & 60 & 12823 & ACIS-S & 1.3 & 154.38 \\
IC\,450 (Mrk\,6) & 12365 & 12365 & WFC3 & FQ508N & 16 & 576 & 210 & 10324 & ACIS-S & 24 & 75.03 \\
IC\,5063 & 8598 & 8598 & WFPC2 & FR533N & 18 & 600 & 80 & 7878 & ACIS-S & 11 & 34.54 \\
M31 (NGC\,224) & 12174 & 12174 & WFC3 & F502N & 0.15 & 2700 & 2700 & 14196 & ACIS-S & 0.23 & 53.83 \\
M51a (NGC\,5194) & 5123 & 5123 & WFPC2/PC & F502N & 1.7 & 1700 & 860 & 13814 & ACIS-S & 2.3 & 192.36 \\
M77 (NGC\,1068) & 5754 & 5754 & WFPC2/PC & F502N & 2.8 & 900 & 440 & 344 & ACIS-S & 3.8 & 48.05 \\
M81 (NGC\,3031) & 1038 & 1038 & WFPC/PC & F502N & 0.79 & 1800 & 350 & 5948 & ACIS-S & 1.1 & 12.18 \\
M90 (NGC\,4569) & 6436 & 6436 & WFPC2/PC & F502N & 2.7 & 1000 & 726 & 5911 & ACIS-S & 3.7 & 39.65 \\
M105 (NGC\,3379) & 6731 & 6731 & WFPC2/PC & F502N & 2.4 & 9785 & 600 & 7073 & ACIS-S & 3.2 & 85.18 \\
M106 (NGC\,4258) & 5123 & 5123 & WFPC2/PC & F502N & 1.7 & 2300 & 1160 & 1618 & ACIS-S & 2.2 & 21.30 \\
NGC\,404 & 12611 & 12611 & WFC3 & F502N & 0.78 & 220 & 400 & 12239 & ACIS-S & 1.2 & 99.56 \\
NGC\,931 & 8598 & 8598 & WFPC2 & FR533N & 24 & 800 & 80 & 12870 & ACIS-S & 15 & 5.03 \\
NGC\,1052 & 3855 & 3855 & WFPC/PC & F502N & 4.2 & 1800 & 350 & 5910 & ACIS-S & 5.9 & 59.96 \\
NGC\,1365 & 6149 & 5222 & WFPC2 & FR533N & 8.8 & 2400 & 220 & 6869 & ACIS-S & 5.4 & 16.12 \\
NGC\,1386 & 6419 & 7278 & WFPC2/PC & F502N & 3.6 & 800 & 280 & 13257 & ACIS-S & 4.8 & 34.26 \\
NGC\,1569 & 8133 & 8133 & WFPC2 & F502N & 1.4 & 1500 & 60 & 782 & ACIS-S & 0.86 & 98.00 \\
NGC\,3393 & 12365 & 12365 & WFC3 & FQ508N & 10.4 & 566 & 208 & 12290 & ACIS-S & 16.0 & 70.05 \\
NGC\,4051 & 12212 & 12212 & WFC3 & F502N & 2.7 & 1554 & 736 & 2148 & ACIS-S & 4.2 & 50.52 \\
NGC\,4151 & 5124 & 5124 & WFPC2/PC & F502N & 2.8 & 870 & 310 & 348 & ACIS-S & 3.8 & 27.95 \\
NGC\,4278 & 6731 & 6731 & WFPC2/PC & F502N & 3.6 & 2600 & 600 & 7081 & ACIS-S & 4.8 & 112.14 \\
NGC\,4388 & 12365 & 12365 & WFC3 & FQ508N & 4.0 & 574 & 194 & 12291 & ACIS-S & 6.2 & 27.96 \\
NGC\,4395 & 12212 & 12212 & WFC3 & F502N & 0.87 & 1521 & 714 & 5302 & ACIS-S & 1.3 & 31.12 \\
NGC\,4507 & 8259 & 8259 & WFPC2 & FR533N & 25 & 520 & 200 & 12292 & ACIS-S & 15 & 43.66 \\
NGC\,5273 & 6419 & 5381 & WFPC2 & FR533N & 7.7 & 400 & 460 & 415 & ACIS-S & 4.8 & 1.76 \\
NGC\,5347 & 8598 & 8598 & WFPC2 & FR533N & 13 & 800 & 80 & 4867 & ACIS-S & 8.1 & 37.41 \\
NGC\,5643 & 5411 & 5411 & WFPC2/PC & F502N & 3.8 & 700 & 50 & 5636 & ACIS-S & 5.0 & 7.94 \\
NGC\,7465 & 6419 & 6419 & WFPC2 & FR533N & 13 & 400 & 40 & 14904 & ACIS-S & 8.1 & 30.06 \\
\hline
\end{tabular}
\end{table}
\end{landscape}

\subsection{\emph{HST} and \emph{Chandra} alignment}
\label{sec:hst_chandra}

The main scope of this work is to study and compare the morphologies of the soft X-ray and [\ion{O}{iii}] emissions. The first caveat to have in mind is the wide differences in the relative astrometry of \emph{Chandra} and \emph{HST} satellites. \emph{Chandra} has an excellent absolute astrometric accuracy of 0.16\arcsec \citep{2011ApJS..192....8R}. However, \emph{HST} is far from this value. Thus, we needed to align both data to ensure a proper analysis. The best alignment would be reached using point-like sources present at both optical and X-ray observations. Unfortunately, this was difficult in our case for two reasons: (1) the small field of view of both \emph{Chandra} and \emph{HST} images strongly reduces the probabilities of finding stars in both fields, and (2) the completely different emission mechanism producing optical and X-ray emission makes rather difficult that the same point-like source is bright enough at both wavelengths. Indeed, this was not possible in our sample. Alternatively, we were able to use the AGN to make the alignment since this source is expected to be bright at both wavelengths. We aligned using the peak in the hard X-ray image, where it is expected to be located the nucleus of the galaxy, with the pixel with the highest flux in the optical image that resemble the nucleus. In some cases we found pile-up in the hard X-ray nuclear source as described in Section~\ref{sec:chandra} (M77, NGC\,3031, and NGC\,4151). This is seen at hard X-rays by a reduction in the number of counts in the affected pixels. For these objects we used the center of symmetry of the nuclear structure. In the case of NGC\,1569 we did not find neither an AGN feature nor a point-like source to align with, so we left it unaltered.

\section{Methodology}
\label{sec:method}

\subsection{Source extension}
\label{sec:source_psf}

To evaluate the optical extent of the sources, firstly we needed to study the point spread function (PSF). The \emph{HST} PSF is similar on the various instruments that we employed in our analysis and it is very stable along the chip. According to the handbooks, 90\% of the PSF encircled energy fraction is found in a radius of $\sim0.3\arcsec$ for the \emph{HST} instruments used in this work. After this, we measured the total flux of the source using a very large aperture different for each object, ensuring that enclosed all the flux by visual inspection. Then, we made aperture photometry with growing radius until reaching a flux that accounts for 90\% of the total flux. If this value was greater than $\sim0.3\arcsec$, the expected for a point source according with the PSF extent, then the source was classified as having an extended compontent. We defined the actual size of the optical emission as this radius accounting the 90\% of the total flux. The rightness of this method relies on the fact that is independent of the actual morphology of the source. It is the best way to establish a size for objects with very different morphologies.

In the case of the \emph{Chandra} data, we evaluated the PSF of the soft and hard X-ray bands for each observation. We performed this analysis using the CIAO tool {\sc psf}\footnote{http://cxc.harvard.edu/ciao/ahelp/psf.html}. The PSF depends on the offset to the central position of the detector, the exposure time, and the energy. Thus, we computed the PSFs for each observation in our sample. However, the PSFs obtained were similar in all the sources. We detected an improvement of the PSF behaviour around 6\,keV, as described in the specification of the instrument, and a slight change on the PSF radius, but the general behaviour was very similar. The 90\% of the enclosed counts were found within a radius that ranges from $(0.76\pm0.17)\arcsec$ at 0.2\,keV, $(1.12\pm0.14)\arcsec$ at 2.0\,keV, and $(5.65\pm0.16)\arcsec$ at 10.0\,keV. We then obtained the light profile analysis of our final images at soft and hard X-rays for each object, centred at the object and extending up to 100 pixels ($\sim6\arcsec$). We calculated the radius that accounts for the 90\% of the flux. We compared this radius with the theoretical value computed from the PSF analysis.

It is important to mention that \emph{Chandra} images have lower spatial resolution than the \emph{HST} ones, implying that \emph{HST} images can host structures that can not be resolved with \emph{Chandra}. Thus, the morphological comparison presented in this paper is based on diffuse extended structures seen in both images.

\subsection{{[O\,III]} luminosities}
\label{sec:oiii_lum}

We computed the [\ion{O}{iii}] luminosities for our sources both in the nuclear region and in the entire area with [\ion{O}{iii}] emission. We defined the nuclear [\ion{O}{iii}] region as that encircled in the PSF size for each object. To calculate this luminosity we performed aperture photometry within the PSF size. In the case of M31, M105, NGC\,1569, and NGC\,4278 we did not find a nuclear feature in the [\ion{O}{iii}] images, then we did not calculate a nuclear [\ion{O}{iii}] luminosity for these objects. In addition, we obtained the total [\ion{O}{iii}] luminosity, as that comprising the nuclear and the extended emission regions. In order to do this, we defined the aperture as the size of the source in Section~\ref{sec:source_psf}. The luminosities for each object are shown in Cols.~6 and 7 in Table~\ref{tab:sample}. The values are not extinction-corrected since we lack of extinction maps to properly correct our [\ion{O}{iii}] flux calibrated images.

\subsection{Comparison images}
\label{sec:comp_img}

Different types of comparison images gave us information about different components of the studied source.

\begin{itemize}
\item[$\rm{\bullet}$]{\underline{[\ion{O}{iii}] and soft X-ray}}: Comparison of pure [\ion{O}{iii}] emission, classical NLR tracer, with the soft X-ray emission.
\item[$\rm{\bullet}$]{\underline{[\ion{O}{iii}]+continuum and soft X-ray}}: Comparison of [\ion{O}{iii}] before continuum subtraction and soft X-ray morphologies.
\item[$\rm{\bullet}$]{\underline{Continuum and soft X-ray}}: Confirmation that the morphological resemblance comes from [\ion{O}{iii}] and not from continuum.
\item[$\rm{\bullet}$]{\underline{Continuum and [\ion{O}{iii}]}}: Global structure of the galaxy and both [\ion{O}{iii}] emission, as a NLR tracer, and star-forming regions.
\item[$\rm{\bullet}$]{\underline{[\ion{O}{iii}]+continuum and [\ion{O}{iii}]}}: Differences from pure [\ion{O}{iii}] and contaminated [\ion{O}{iii}] emission.
\item[$\rm{\bullet}$]{\underline{Soft X-ray and hard X-ray}}: Extended X-ray emission and nuclear AGN regions.
\end{itemize}

\begin{figure*}
\includegraphics[width=0.66\columnwidth]{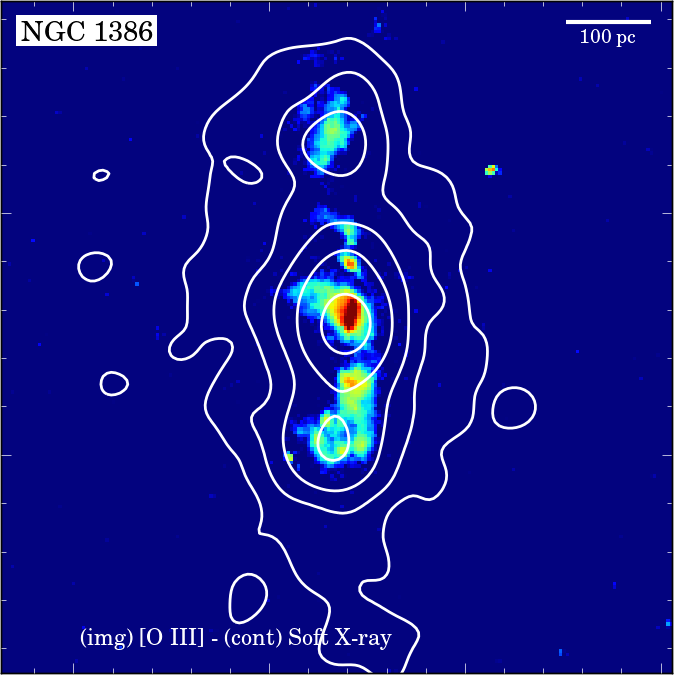}
\vspace{0.1cm}
\hspace{0.1cm}
\includegraphics[width=0.66\columnwidth]{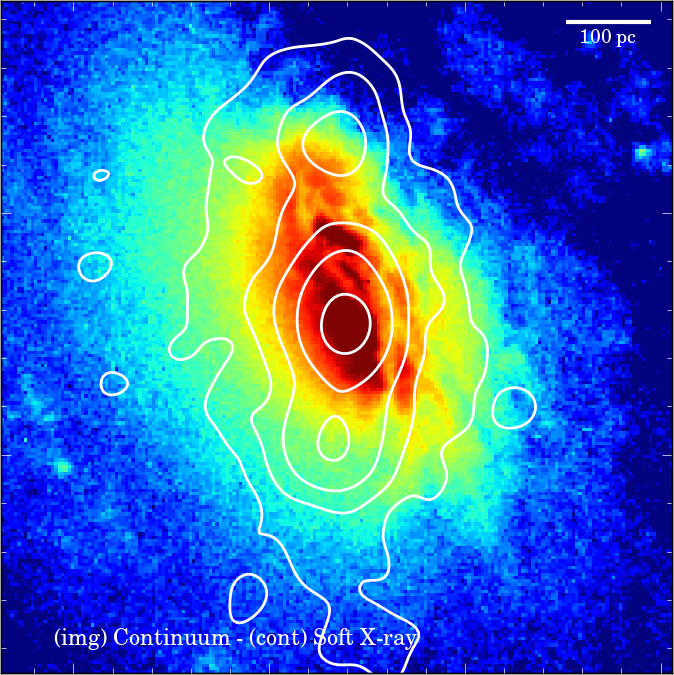}
\vspace{0.1cm}
\hspace{0.1cm}
\includegraphics[width=0.66\columnwidth]{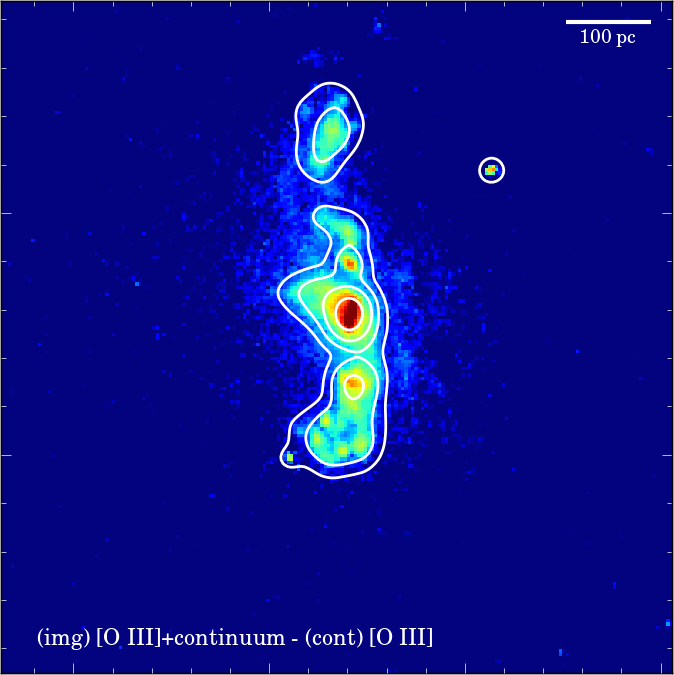}
\vspace{0.1cm}
\includegraphics[width=0.66\columnwidth]{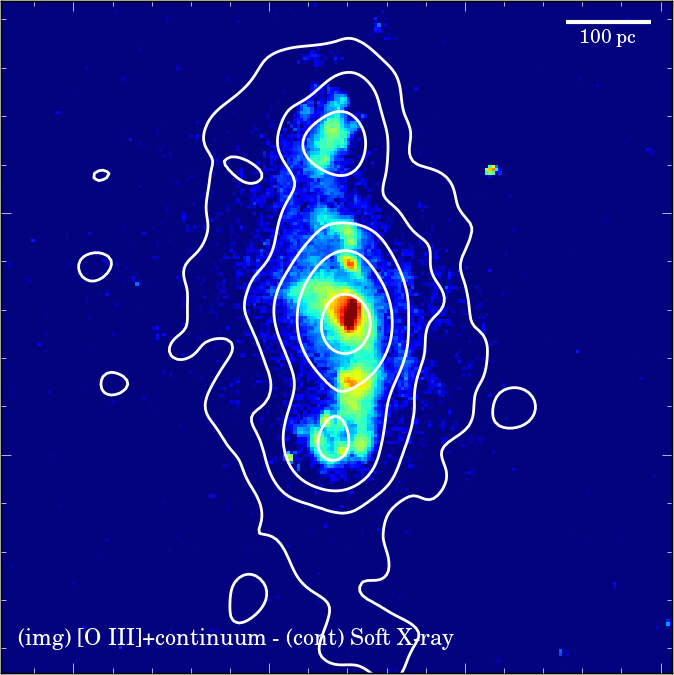}
\hspace{0.1cm}
\includegraphics[width=0.66\columnwidth]{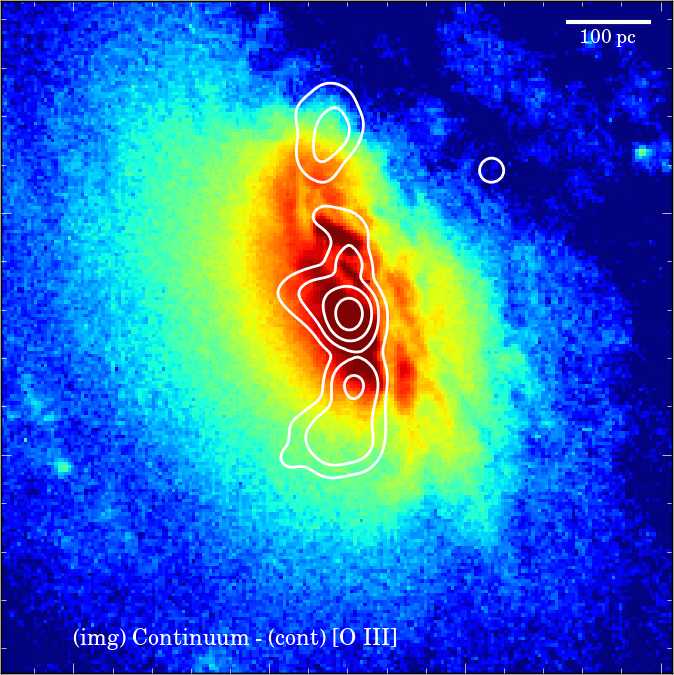}
\hspace{0.1cm}
\includegraphics[width=0.66\columnwidth]{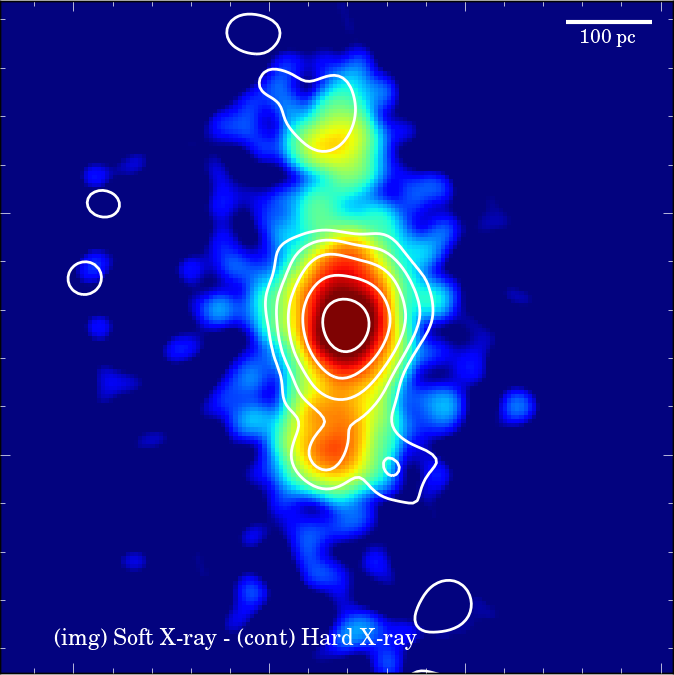}
\caption{Set of images for NGC\,1386 (see text). North is up, East is to the left. \emph{Top left panel}: \emph{Chandra} soft X-ray contours superimposed on \emph{HST} [\ion{O}{iii}] image. \emph{Bottom left panel}: \emph{Chandra} soft X-ray contours superimposed on \emph{HST} [\ion{O}{iii}]+continuum image. \emph{Top middle panel}: \emph{Chandra} soft X-ray contours superimposed on \emph{HST} continuum image. \emph{Bottom middle panel}: \emph{HST} [\ion{O}{iii}] contours superimposed on \emph{HST} continuum image. \emph{Top right panel}: \emph{HST} [\ion{O}{iii}] contours superimposed on \emph{HST} [\ion{O}{iii}]+continuum image. \emph{Bottom right panel}: \emph{Chandra} hard X-ray contours superimposed on \emph{Chandra} soft X-ray image. The contours show logarithmic intervals with the minimum value set to 3$\sigma$ over the background and the maximum set to 50\% of the peak value. The images are scaled in the same way. We apply a smooth value of 3 pixels to plot the contours in all the images. The center of the images correspond to the soft X-ray peak. Note that the full set of \emph{Chandra} soft X-ray contours superimposed on \emph{HST} [\ion{O}{iii}] images is included in Appendix \ref{sec:appendix}.}
\label{fig:comp_ngc1386}
\end{figure*}

We compared soft X-rays with the [\ion{O}{iii}] emission line before and after the continuum subtraction and with the BB image in order to understand to which of them the soft X-ray image resembles better. We also compared soft with hard X-rays to understand whether the extended emission contributes only to the soft X-rays. In Fig.~\ref{fig:comp_ngc1386}, we include the full set of image comparisons for the case of NGC\,1386 as an example. In this case, it is interesting to note that the soft X-ray emission resembles the [\ion{O}{iii}] emission both before and after continuum subtraction. However, this soft X-ray emission does not resemble the continuum emission of the host.
We further investigate this aspect for the full sample below. In Appendix~\ref{sec:appendix}, the soft X-ray and [\ion{O}{iii}] comparison images for the 27 AGN in our sample are included.

\section{Results}
\label{sec:results}

\subsection{X-ray morphologies}
\label{sec:xray_morph}

In order to decide whether a source is extended or not we compared the \emph{Chandra} PSF radius with the radius of the soft and hard X-ray bands emission using the procedure described in Section~\ref{sec:source_psf}. In all cases, the soft X-ray sources detected in our sample show radii larger than that of the PSF, confirming that all the objects are extended in the soft X-ray regime, regardless of the AGN class. On the other hand, at hard X-rays the majority of the sources in our sample are compact, except for four objects (M51a, M77, NGC\,1386, and NGC\,5643). In these cases the soft X-ray morphology is more extended than that in the hard X-rays, but the hard X-ray images show similar morphology compared to the brightest structures seen at soft X-rays. All of them are type-2 Seyferts.

We performed a morphological classification of the sources for both soft and hard X-ray energy regimes. In the soft X-ray regime all the objects show diffuse emission as mentioned above, but we separated them between `cone-shaped', `spheroidal', or `other' morphologies, that are typically diffuse emission following the general morphology of the galaxy. Among the hard X-ray emission, we distinguished between diffuse and point-like morphologies as studied above. The results are shown in Table~\ref{tab:xray_morph} (see Col.~7 in Table~\ref{tab:sample} for the classification of each source).

\begin{table}
\caption{Summary of the X-ray morphologies. `Cone' denotes cone-shaped, `Sph.' stands for spheroidal, `Other' suggests diffuse morphologies that can not be classified into cone-shaped or spheroidal, `Diff.' indicates diffuse, and `Point' denotes point-like sources. We show the percentages in each cell in brackets, calculated over the number of sources within each optical type.}
\label{tab:xray_morph}
\begin{tabular}{l c c c c c}
\hline
Optical & \multicolumn{3}{c}{Soft X-ray} & \multicolumn{2}{c}{Hard X-ray} \\
type & Cone & Sph. & Other & Diff. & Point \\
\hline
Sy1 (9) & 2 & 6 & 1 & 0 & 9 \\
 & (22\%) & (67\%) & (11\%) & (0\%) & (100\%) \\
Sy2 (10) & 8 & 0 & 2 & 3 & 7 \\
 & (80\%) & (0\%) & (20\%) & (30\%) & (70\%) \\
LINER (8) & 4 & 4 & 0 & 0 & 8 \\
 & (50\%) & (50\%) & (0\%) & (0\%) & (100\%) \\
\hline
\end{tabular}
\end{table}

We can see that in the case of the type-1 Seyferts in our sample, the dominant morphology in the soft X-ray energy regime is spheroidal. In the hard X-rays, all the objects show point-like structures. This situation change in the type-2 Seyferts, mainly displaying cone-shaped emission in the soft X-rays and it appears a relevant contribution of diffuse morphologies in the hard X-rays. The LINERs in our sample do not present a dominant morphology in the soft X-rays, being both cone-shaped or spheroidal, and in the hard X-rays they are point-like sources.

Among the 10 type-2 Seyferts presented here, six are in common with the sample studied by \citet{2006A&A...448..499B}, namely NGC\,1386, NGC\,3393, NGC\,4388, NGC\,4507, NGC\,5347, and NGC\,5643. \citet{2006A&A...448..499B} found extended emission in the soft X-ray regime for all the sources in their eight type-2 Seyfert sample, except for the case of NGC\,4507, severely affected by pile-up. For this source we were able to detect extended emission in this band. This is because we chose a more recent observation not affected by pile-up, which was not available by the time \citet{2006A&A...448..499B} presented their analysis. In the case of the hard X-ray regime, \citet{2006A&A...448..499B} did not find any evidence for extended emission beyond the PSF in their sources. We detected extended emission in four objects in this band, among them NGC\,1386 and NGC\,5643 shared with \citet{2006A&A...448..499B} sample. In the case of NGC\,1386, the set of \emph{Chandra} data that we used in this work is more recent and has a greater exposure time. For NGC\,5643, we employed the same data, so the disagreement could come from the different reduction techniques. We employed smoothing techniques that allow us to detect the low-contrast diffuse emission, something not mentioned in \citet{2006A&A...448..499B} reduction method.

\subsection{{[O\,III]} morphologies}
\label{sec:oiii_morph}

In the optical, all the sources in our sample present diffuse emission beyond the \emph{HST} PSF. Note that the type-1 Seyferts IC\,450, NGC\,4051, and NGC\,4395 are saturated in the central region, but this does not affect the external features we are interested in.

We calculated the contribution of the [\ion{O}{iii}] extended emission over the total [\ion{O}{iii}] luminosity for each optical type. All the objects show an important contribution of the extended emission, at least a 20\%. In the type-1 Seyferts the scatter is large, but the sources cluster around two regimes where the dispersion reduces significantly, above and below a 50\%. In type-2 Seyferts and LINERs the [\ion{O}{iii}] luminosity is dominated by the extended emission, with only one object of each class below 50\% (M90 and NGC\,4507). Table~\ref{tab:lum_ext} shows the average and standard deviation within each optical type and regime above and below a 50\%.

\begin{table}
\caption{Contribution of the extended [\ion{O}{iii}] luminosity over the total [\ion{O}{iii}] luminosity. In Col.~1 we show the average and standard deviation considering all the objects within each optical type. In Cols.~2 and 3 we only count those objects with its extended contribution above and below a 50\% respectively.}
\label{tab:lum_ext}
\centering
\begin{tabular}{l c c c}
\hline\hline
Opt. type & All & $>50\%$ & $<50\%$ \\
\hline
Sy1 & 0.56 $\pm$ 0.30 & 0.84 $\pm$ 0.10 & 0.29 $\pm$ 0.02 \\
Sy2 & 0.85 $\pm$ 0.16 & 0.89 $\pm$ 0.09 & 0.47 \\
LINER & 0.84 $\pm$ 0.22 & 0.93 $\pm$ 0.09 & 0.47 \\
\hline
\end{tabular}
\end{table}

We classified the morphology of the sources in the same way as we did for the X-rays. The results within each optical type are gathered in Table~\ref{tab:opt_morph} (see Col.~8 in Table~\ref{tab:sample} for the classification of each source).

\begin{table}
\caption{Summary of the [\ion{O}{iii}] morphologies. `Cone' denotes cone-shaped, `Sph.' stands for spheroidal, `Other' suggests diffuse morphologies that can not be classified into cone-shaped or spheroidal. We show the percentages in each cell in brackets, calculated over the number of sources within each optical type.}
\label{tab:opt_morph}
\centering
\begin{tabular}{l c c c}
\hline\hline
Opt. type & Cone & Sph. & Other \\
\hline
Sy1 (9) & 4(45\%) & 3(33\%) & 2(22\%) \\
Sy2 (10) & 9(90\%) & 0(0\%) & 1(10\%) \\
LINER (8) & 0(0\%) & 7(88\%) & 1(12\%) \\
\hline
\end{tabular}
\end{table}

In the type-1 Seyferts in our sample, the numbers do not show a significant difference between spheroidal and cone-shaped. We do not find a dominant morphology. The type-2 Seyfert objects present predominantly cone-shaped morphologies, none of them displaying spheroidal structures. For the LINERs, it is very representative the total absence of cone-shaped structures in favour of a spheroidal appearance. Although our sample is not complete, we would like to stress that this result is in agreement with the previous work done by \citet{2003ApJ...597..768S}, whose morphological study of [\ion{O}{iii}]-traced NLRs in Seyfert galaxies revealed that there was a higher percentage of type-1 Seyferts with halo-like NLRs, while those in type-2 Seyferts were more elongated.

\begin{figure}
\resizebox{\hsize}{!}{\includegraphics{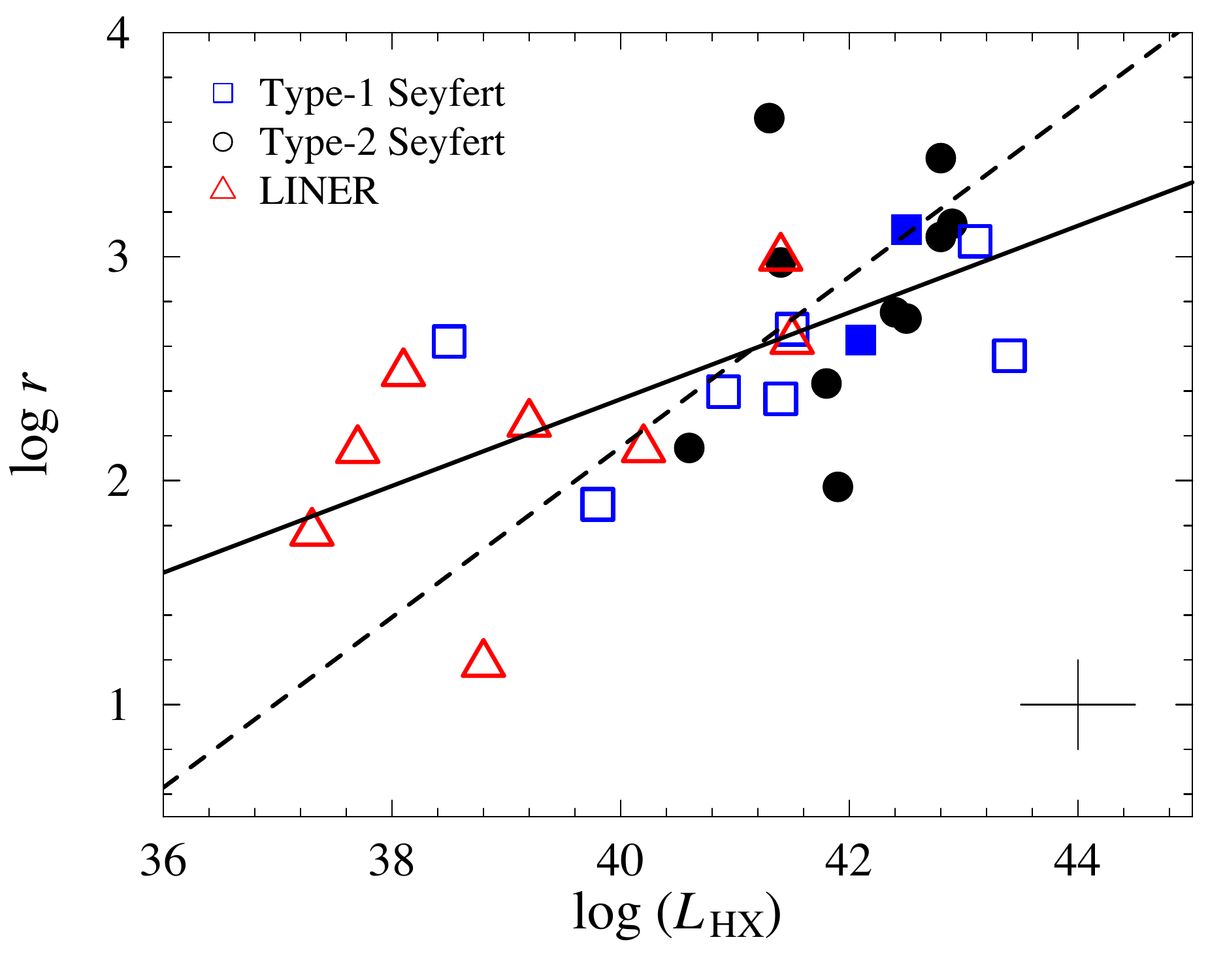}}
\caption{Logarithm of the radius in pc versus the logarithm of the intrinsic hard X-ray luminosity in CGS units. The filled symbols refer to those objects that show a good match between the soft X-ray and the [\ion{O}{iii}] morphologies. The solid line shows the best linear fit to all the galaxies. The dashed line shows \citet{2011A&A...527A..23M} best linear fit to their `core-halo' sample of LINERs. Average error bars are in the bottom right corner.}
\label{fig:hard_r}
\end{figure}

If [\ion{O}{iii}] is tracing the photoionised extended NLR, we would expect a tight correlation between the size of the [\ion{O}{iii}] emitting region and the power of the AGN, traced by the hard X-ray luminosity. To further investigate this, Fig.~\ref{fig:hard_r} shows the relation between the size of the [\ion{O}{iii}] emission and the hard X-ray luminosity. This relation was previously explored for the NLR of Seyfert galaxies \citep{2002ApJ...574L.105B,2003ApJ...597..768S} using the [\ion{O}{iii}] luminosity as a proxy of the AGN power. Later on, \citet{2011A&A...527A..23M} studied this relation for a sample of LINERs employing the hard X-ray luminosity as a better way to trace the AGN power \citep{2003MNRAS.344L..59M}. We followed the last approach. We calculated the radius of the size of the [\ion{O}{iii}] emission for each object as that enclosing the total [\ion{O}{iii}] luminosity (see Sect.~\ref{sec:oiii_lum}). Spearman and Kendall rank correlation tests led to $p\rm{-values}<10^{-3}$. Our slope value is $0.19\pm0.05$ with a correlation coefficient of 0.630, although the dispersion is high. This slope is lower than the results obtained by \citet{2003ApJ...597..768S} for Seyfert galaxies ($0.33\pm0.04$, $r=0.627$) and by \citet{2011A&A...527A..23M} for its `core-halo' sample of LINERs ($0.38\pm0.05$, $r=0.949$). \citet{2011A&A...527A..23M} employed H$\alpha$ in their analysis and not [\ion{O}{iii}], thus, we could be tracing a different mechanism.

Another way to quantify the relation between the [\ion{O}{iii}] emission and the AGN power, again expected if [\ion{O}{iii}] traces the NLR of the AGN, is to study the correlation between the hard X-ray and [\ion{O}{iii}] emissions, as done by \citet{2006A&A...455..173P} spectroscopically. Since the hard X-rays are a tracer of the AGN power, then such a correlation implies that also [\ion{O}{iii}] is a proxy for photoionisation. Finding morphological similarities between the soft X-ray and [\ion{O}{iii}] emissions would imply that the soft X-rays are also a tracer of photoionisation. In order to study the hard X-ray and [\ion{O}{iii}] correlation for our sample, corrections from optical extinction in our [\ion{O}{iii}] luminosities measurements would be needed. Unfortunately, we lack of extinction maps to properly account for the correction, and thus we refrained from examining the correlation.

\subsection{Soft X-ray versus {[O\,III]} morphologies}
\label{sec:soft_oiii_morph}

We studied the resemblance between the soft X-ray and [\ion{O}{iii}] morphologies. Making use of [\ion{O}{iii}] emission-line images taken with \emph{HST} and X-ray images from \emph{Chandra} we ensure a comparison at the best spatial resolution that is accessible with the current instrumentation. The high angular resolution of these satellites is the best to trace the NLR morphology.

We found that the soft X-ray emission resembles the overall morphology of the continuum-subtracted [\ion{O}{iii}] emission in 12 out of the 27 objects. In Table~\ref{tab:results}, we summarise these results. None of the spheroidal morphologies found at X-rays match with the [\ion{O}{iii}] morphologies. Furthermore, we only detect a good morphological correlation between the soft X-ray and the [\ion{O}{iii}] emissions in objects with high intrinsic hard X-ray luminosities ($\log {(L_{\rm{HX}})}>40$). The matching is total in the case of type-2 Seyferts (100\%), relatively rare for type-1 Seyferts (22\%), and absent in all LINERs (see Col.~7 in Table~\ref{tab:sample} for the match on each individual source).

Among the type-2 Seyfert galaxies, Circinus, M51a, and M77 do not present a good match in terms of extension of both types of emission. For these cases the soft X-ray emission appears widely more extended than the [\ion{O}{iii}], but just for the low brightness contours. The resemblance is clear in the structures with a high S/N level. In all the type-2 Seyferts we can see that the soft X-ray emission reaches lower brightness regions than the [\ion{O}{iii}] images. This is also the case in the type-1 Seyfert NGC\,4151, but not in NGC\,1365, both presenting a good soft X-ray/[\ion{O}{iii}] match. For the rest type-1 Seyferts, in which both emissions do not resemble, the [\ion{O}{iii}] is more extended than the soft X-rays in four cases (M106, NGC\,1569, NGC\,4051, and NGC\,4395), along with NGC\,1365, a 56\% of the total type-1 Seyferts sample. This number is 1/8 (12\%) for the LINERs.

\begin{table}
\caption{Summary of the matches between soft X-ray and [\ion{O}{iii}] morphologies.}
\label{tab:results}
\centering
\begin{tabular}{l c c c}
\hline\hline
Optical type & Match & Total & Percentage \\
\hline
Sy1 & 2 & 9 & 22\% \\
Sy2 & 10 & 10 & 100\% \\
LINER & 0 & 8 & 0\% \\
\hline
\end{tabular}
\end{table}

\begin{figure*}
\begin{center}
\includegraphics[width=0.66\columnwidth]{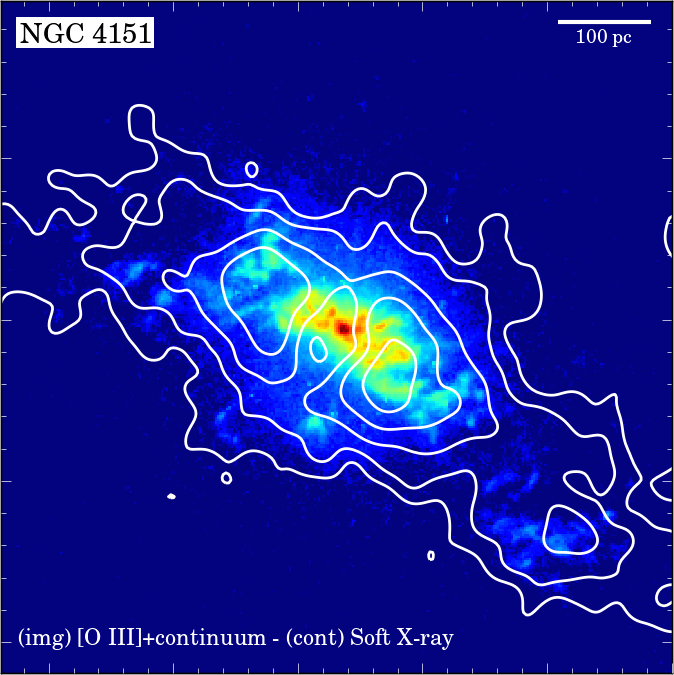}
\vspace{0.1cm}
\hspace{0.1cm}
\includegraphics[width=0.66\columnwidth]{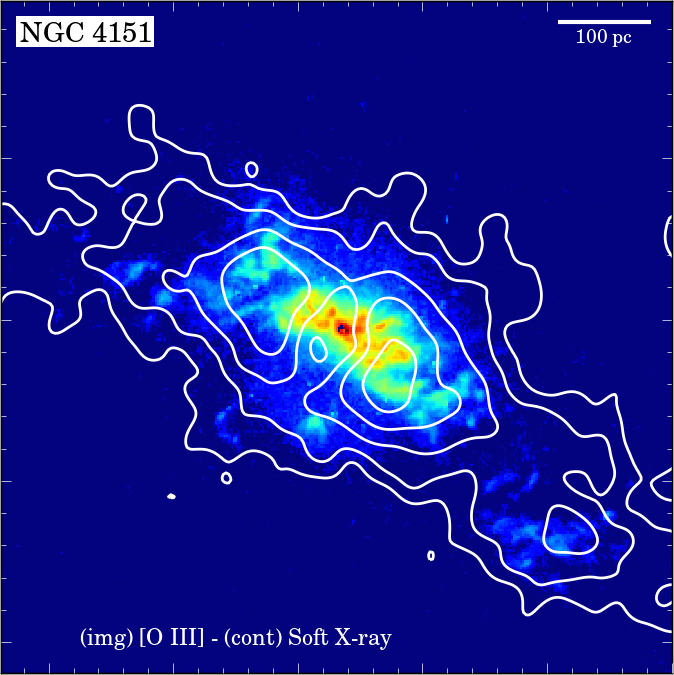}\\
\vspace{0.1cm}
\includegraphics[width=0.66\columnwidth]{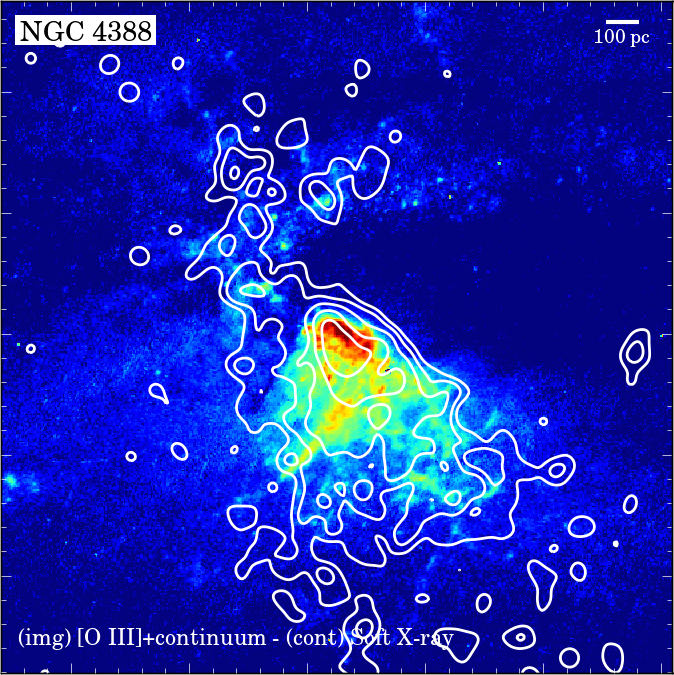}
\hspace{0.1cm}
\includegraphics[width=0.66\columnwidth]{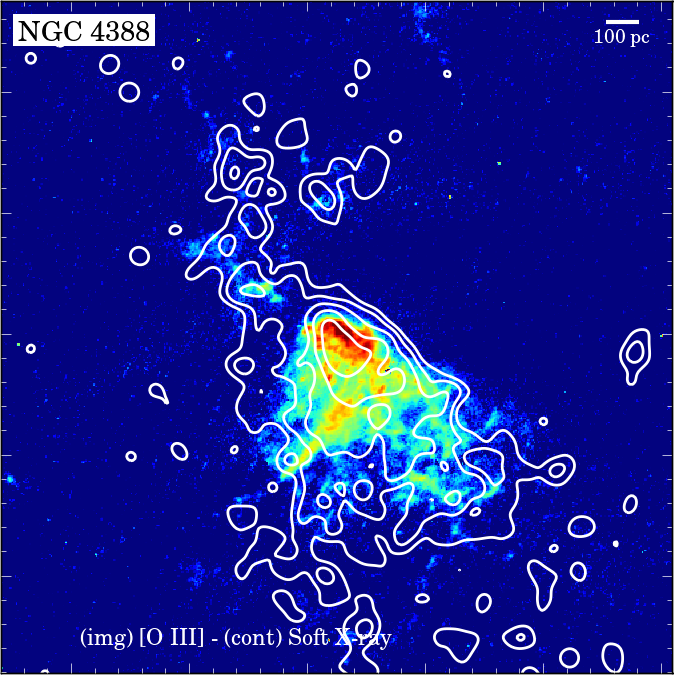}
\caption{Effect of continuum subtraction for NGC\,4151 and NGC\,4388 (see text). North is up, East is to the left. \emph{Top left panel}: NGC\,4151 before continuum subtraction. \emph{Chandra} soft X-ray contours superimposed on \emph{HST} [\ion{O}{iii}]+continuum image. \emph{Top right panel}: NGC\,4151 after continuum subtraction. \emph{Chandra} soft X-ray contours superimposed on \emph{HST} [\ion{O}{iii}] image. \emph{Bottom left panel}: NGC\,4388 before continuum subtraction. \emph{Chandra} soft X-ray contours superimposed on \emph{HST} [\ion{O}{iii}]+continuum image. \emph{Bottom right panel}: NGC\,4388 after continuum subtraction. \emph{Chandra} soft X-ray contours superimposed on \emph{HST} [\ion{O}{iii}] image.}
\label{fig:subs_ngc4151_ngc4388}
\end{center}
\end{figure*}

We obtain exactly the same results comparing soft X-ray and [\ion{O}{iii}]+continuum images. All the sources showing a good match between soft X-ray and [\ion{O}{iii}] images still show that resemblance with the [\ion{O}{iii}] image prior continuum subtraction. For example, the effect of continuum subtraction in NGC\,4151 is shown in Fig.~\ref{fig:subs_ngc4151_ngc4388}. The difference between both non-subtracted and subtracted images is minimum. Then, continuum subtraction seems not to be crucial in the analysis, indicating that the [\ion{O}{iii}] emission line fully dominates the filter. However, a proper continuum subtraction is important in order to get rid of features that might confuse the correlation in some sources. For instance, this is the case of the type-2 Seyfert NGC\,4388 that is shown in Fig.~\ref{fig:subs_ngc4151_ngc4388}, where the dust band in the galaxy plane does not correlate with the soft X-ray emission, but it disappears with the continuum subtraction allowing a better match between both morphologies.

Our sample contains six type-2 Seyferts in common with \citet{2006A&A...448..499B} for which we find the same morphological correlation. These authors reported a good match between the soft X-ray and [\ion{O}{iii}] emissions for all their eight type-2 Seyfert sample. We include four additional type-2 Seyferts, which means an increment of 40\%. Our results show an universal correlation between the soft X-ray and [\ion{O}{iii}] emissions in type-2 Seyferts, such as that evidenced in \citet{2006A&A...448..499B}. Taking into account all the optical types analysed in our work the correlation between both emissions is much lower, 44\% (12/27).

\begin{figure}
\resizebox{\hsize}{!}{\includegraphics{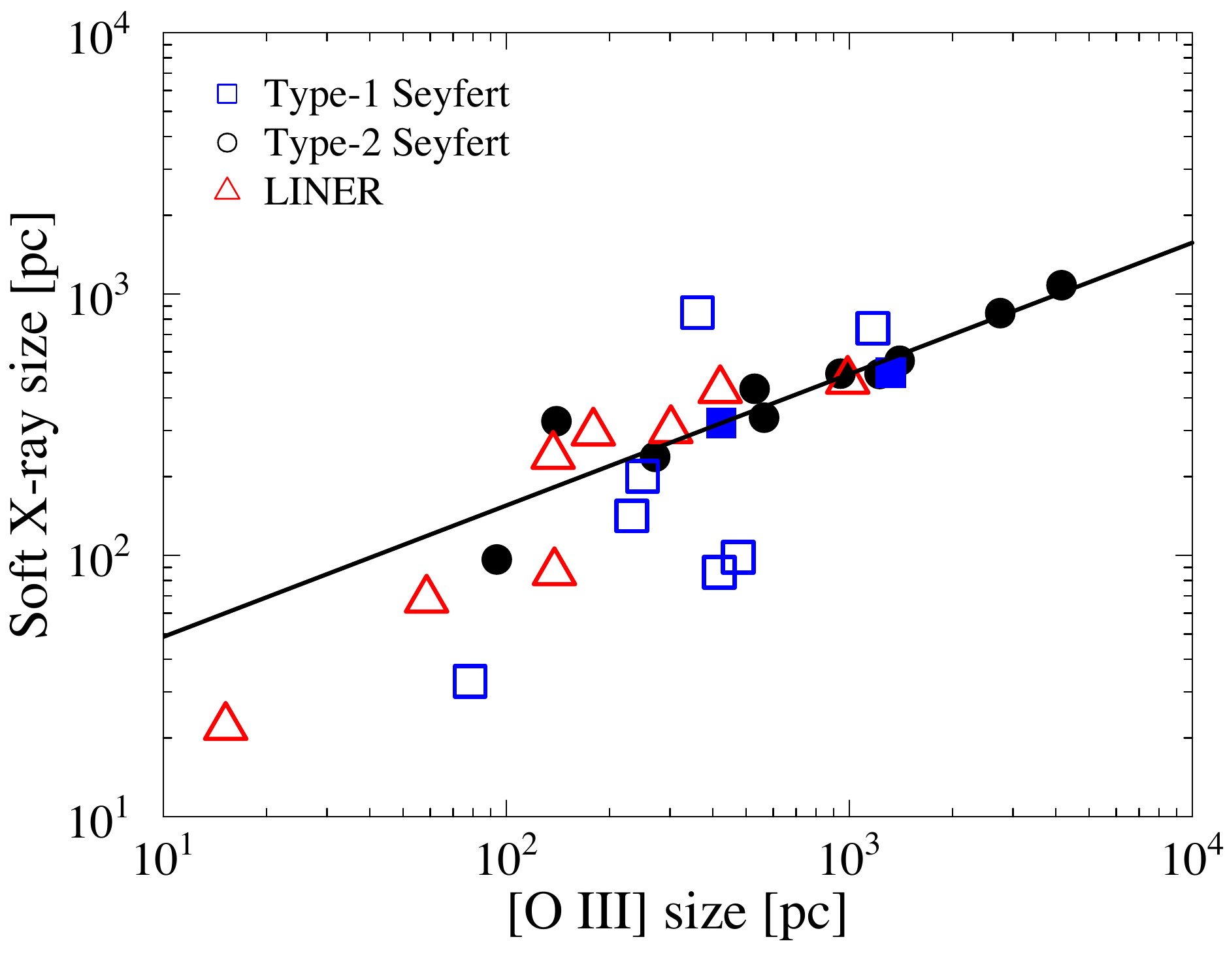}}
\caption{Soft X-ray versus [\ion{O}{iii}] emission region sizes. The filled symbols refer to those objects showing a good match between the soft X-ray and [\ion{O}{iii}] morphologies, along with the best linear fit to them as a solid line.}
\label{fig:comp_sizes}
\end{figure}

The good match between the soft X-ray and [\ion{O}{iii}] morphologies is an argument in favour for a common physical origin of both emissions in the sources showing this correlation. In order to support this, in Fig.~\ref{fig:comp_sizes} we show a comparison of the sizes of these two emissions as defined in Sect.~\ref{sec:source_psf}. We detect a high degree of correlation in sizes for the sources showing a good correspondence in morphologies. Spearman and Kendall rank correlation tests lead to $p\rm{-values}<10^{-3}$, and the correlation coefficient is 0.924.

\section{Discussion}
\label{sec:discussion}

Under the UM, the NLR could be an outflow of material coming from the AGN, launched by the accretion disk and collimated by the torus \citep{2000ApJ...530L..65N}. We expect morphological differences between the NLR of type-1 and type-2 Seyfert galaxies due to the viewing angle. The conical NLR of type-1 Seyferts is observed closer to face-on, while in the case of type-2 Seyferts is observed closer to edge-on. Then, while we expect conical shapes for the NLR of type-2 Seyferts, in the type-1 we would rather find more spheroidal shapes (we refer the reader to \citet{2011ApJ...731...92R} for further discussion on torus inclination). Indeed, under the classical hypothesis that the [\ion{O}{iii}] emission traces the NLR \citep{2003ApJS..148..327S,2003ApJ...597..768S}, we found a higher percentage of spheroidal [\ion{O}{iii}] morphologies in type-1 Seyferts, in agreement with the UM scenario. However, we found four cases (45\%) of conical NLR in our type-1 Seyfert sample. This could be interpreted as a continuous variation in the angle along the LOS respect to the torus. Fully face-on cases would lead to spheroidal morphologies, but less extreme angles would allow partially seeing the BLR, classifying the galaxy as type-1 Seyfert, while allowing the observation of a conical NLR.

In all the type-2 Seyferts we detected a good match between soft X-ray and [\ion{O}{iii}] morphologies. This was already reported by \citet{2006A&A...448..499B}. It would imply that the physical origin for the soft X-ray emission is also AGN photoinisation (the so-called NLR). This was reinforced by the study of the high resolution spectra of type-2 Seyferts presented by \citet{2007MNRAS.374.1290G}. They showed that a strong contribution of photoionised gas by the AGN is needed to explain the emission lines seen at soft X-rays in most of these type-2 Seyferts.

\begin{figure*}
\begin{center}
\includegraphics[width=0.66\columnwidth]{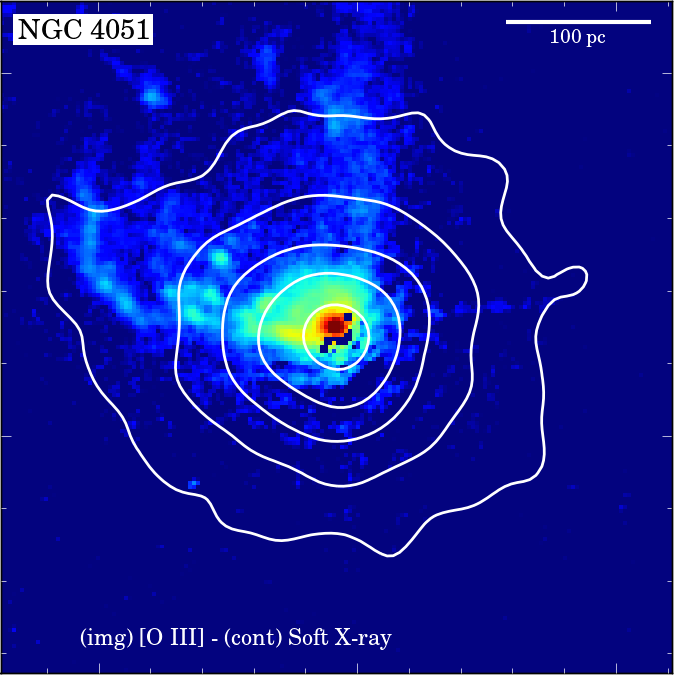}
\hspace{0.1cm}
\includegraphics[width=0.66\columnwidth]{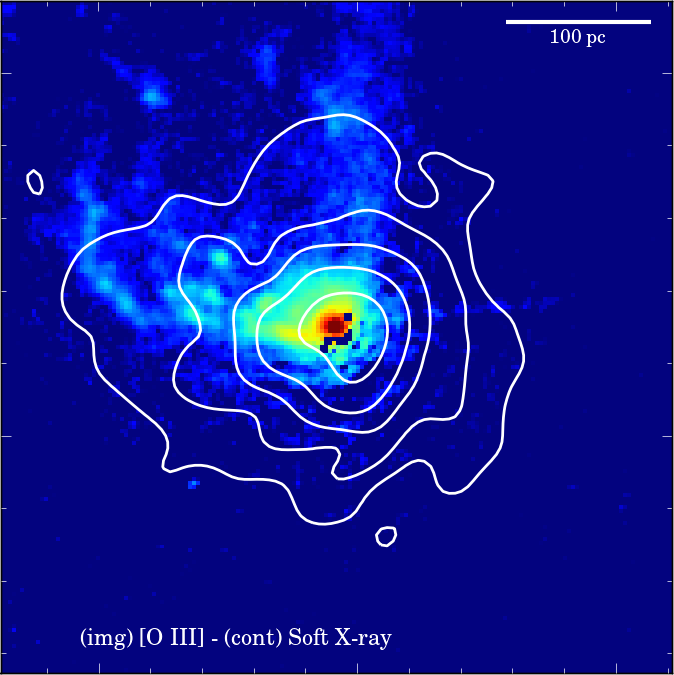}
\caption{Effect of hard X-ray subtraction on soft X-ray/[\ion{O}{iii}] comparison for NGC\,4051 (see text). North is up, East is to the left. \emph{Left panel}: NGC\,4051 before hard X-ray subtraction. \emph{Chandra} soft X-ray contours superimposed on \emph{HST} [\ion{O}{iii}]. \emph{Right panel}: NGC\,4051 after hard X-ray subtraction. \emph{Chandra} soft X-ray contours subtracted from hard X-rays superimposed on \emph{HST} [\ion{O}{iii}] image. Note that none smooth is applied to plot the contours for this case.}
\label{fig:sy1_sub}
\end{center}
\end{figure*}

In the case of type-1 Seyferts, we found that only two out of nine presented a good match between soft X-ray and [\ion{O}{iii}] morphologies (NGC\,1365 and NGC\,4151). A good correlation between both emissions was reported for the archetypical galaxy NGC\,4151 \citep{2000ApJ...545L..81O,2001ApJ...563..124Y}. On the premise that the same mechanism seen in type-2 Seyferts should be in force in the type-1 Seyferts, we also expected a good match between soft X-ray and [\ion{O}{iii}] morphologies. However, in those objects closer to a face-on configuration, the X-ray nuclear emission could outshine the extended emission, preventing a good discrimination. To better explore this, we subtracted the hard X-ray emission, assumed as a PSF, to the soft X-ray image in NGC\,4051. That is, we scaled the nuclear emission to the 90\% of the peak value and subtract it. The results are shown in Fig.~\ref{fig:sy1_sub}. After subtraction, the soft X-ray contours do not show the same spheroidal appearance as before, but they do allow a glimpse of conical morphology. We applied the same procedure to NGC\,4395 and NGC\,5273. Nevertheless, in these cases we do not see any match.

\begin{figure}
\resizebox{\hsize}{!}{\includegraphics{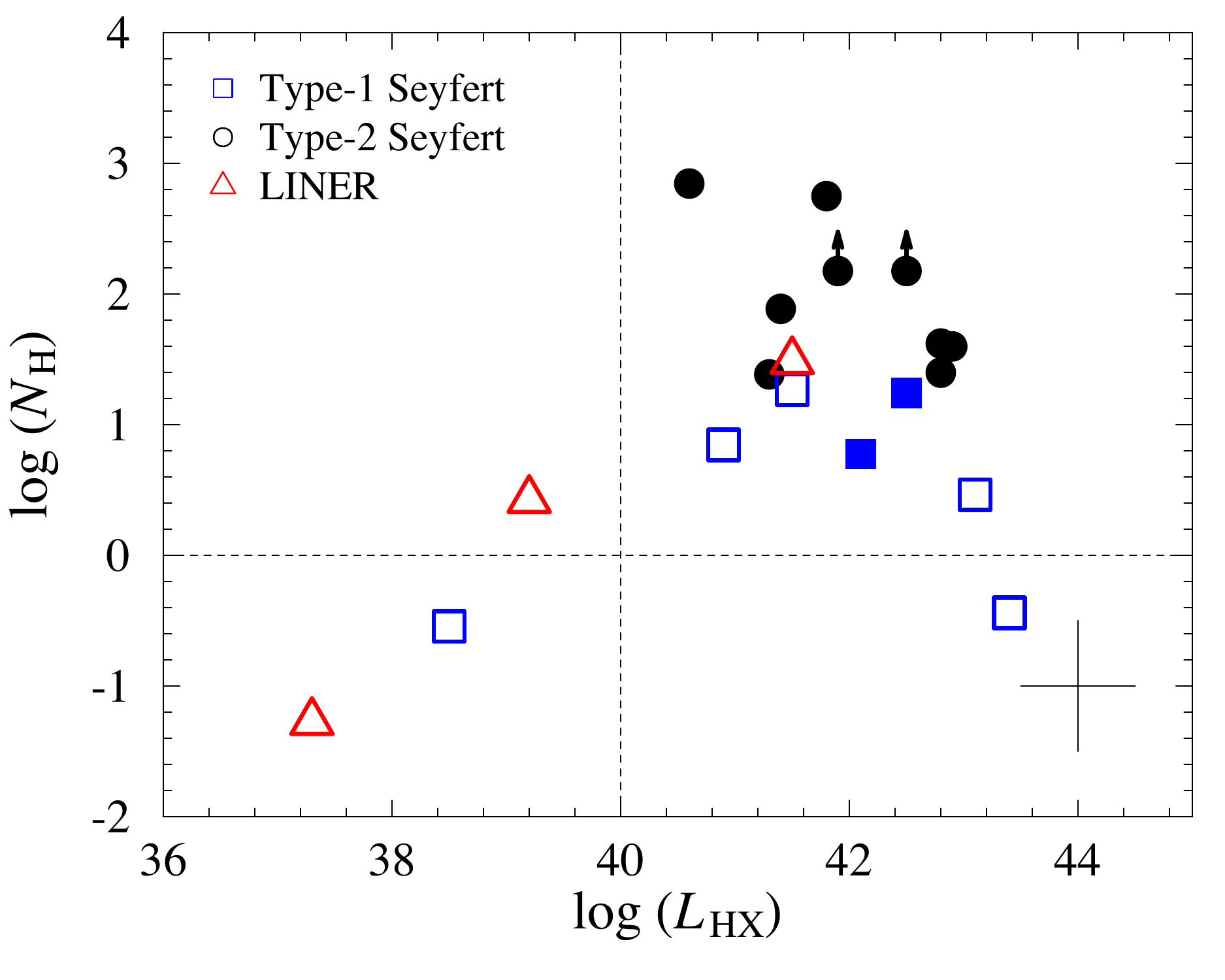}}
\caption{Logarithm of the column density ($10^{22}$\,cm$^{-2}$) versus the logarithm of the intrinsic hard X-ray luminosity in CGS units. The filled symbols refer to those objects that show a good match between the soft X-ray and the [\ion{O}{iii}] morphologies. The dashed lines shows $N_{\rm{H}} = 10^{22}$\,cm$^{-2}$ and $\log(L_\mathrm{HX}) = 40$ as references. Average error bars are in the bottom right corner.}
\label{fig:hard_nh}
\end{figure}

Building on this, it is very important to say that the column density plays a main role in what is observed in the X-rays \citep{1998A&A...338..781M}. In unobscured objects the soft X-rays are dominated by the primary emission, whereas in obscured sources the primary emission is not visible and the soft X-rays are associated with extended regions further away from the nucleus and the obscurer, that can be linked with the NLR. Therefore, it would be expected to see extended soft X-ray emission matched with the NLR as traced by [\ion{O}{iii}] predominantly in objects with high column densities. Indeed, objects are typically classified as unobscured if $N_{\rm{H}} < 10^{22}$\,cm$^{-2}$ and obscured if $N_{\rm{H}} > 10^{22}$\,cm$^{-2}$. In Fig.~\ref{fig:hard_nh}, we plot the column density and the hard X-ray luminosity for those objects with both measurements reported in the references shown in Table~\ref{tab:sample}. We see that column densities a few times above $10^{22}$\,cm$^{-2}$ and hard X-ray luminosities $\log(L_\mathrm{HX}) > 40$ display a good match between soft X-ray and [\ion{O}{iii}] emissions in 11 out of 15 objects (73\%), and all measured sources with such match lie on this parameter space.

It is important to discuss the role of other processes as potential main sources to originate both the soft X-ray and [\ion{O}{iii}] emissions. We have already mentioned in Section~\ref{sec:intro} the case of NGC\,1365, where the diffuse soft X-ray emission can be explained only through a thermal origin \citep{2009ApJ...694..718W}. This object is a case of good match between soft X-ray and [\ion{O}{iii}] morphologies, so it seems plausible that the thermal component is the main mechanism to originate both types of emission in this source. In addition, radio jets can create shocks and heat the ISM. The presence of radio jets is confirmed in most of the type-1 and type-2 objects in our sample, being in most of the cases consistent with the orientation of the soft X-ray and [\ion{O}{iii}] emissions. This is the case of the type-2 Seyferts Circinus \citep{2012ApJ...758...95M}, IC\,5063 \citep{2003ApJS..148..327S}, M51a \citep{2007MNRAS.379.1249D}, NGC\,1386 \citep{2003ApJS..148..327S}, NGC\,3393 \citep{2000ApJS..129..517C}, NGC\,4388 \citep{1998ApJ...502..199F}, and NGC\,5643 \citep{1997ApJ...474..121S}, and the type-1 Seyferts NGC\,1365 \citep{2001ApJS..136...61S} and NGC\,4151 \citep{2007MNRAS.379.1249D}. Under the UM, it seems likely that if the radio jet exists it would be located along the same direction than the NLR. The question that arises is whether the jet shocks are the main cause originating the soft X-ray and [\ion{O}{iii}] morphologies. \citet{2000ApJS..129..517C} and \citet{1997ApJ...474..121S} argued for NGC\,3393 and NGC\,5643, respectively, that the primary source was photoionisation from a central source and not radio shocks. However, \citet{2012ApJ...758...95M} found the opposite explanation to originate the soft X-ray emission in Circinus. Therefore, a thermal or radio shock mechanism can not be ruled out as main sources to originate both the soft X-ray and [\ion{O}{iii}] when it comes to individual objects (e.g., Circinus, NGC\,1365).

With all of this in mind, our last question is: what powers soft X-rays and [\ion{O}{iii}] in LINERs? This has not a clear answer, but both are not tracing the same mechanism, since none of them match in morphologies. This is a clear difference between type-2 Seyferts and LINERs. In favour for the soft X-ray emission being originated by AGN photoionisation, the RGS spectra studied by \citet{2010AIPC.1248..343G} showed that in at least 30\% of their sample a contribution of photoionisation by the AGN is required due to the presence of the CV RRC emission line. However, this does not guarantee a dominance of this emission mechanism. Moreover, cone-like morphologies at soft X-rays in some objects in the present study points out again to the photoionisation by the AGN as the responsible for the soft X-ray emission. Nevertheless, this assumes we are seen LINERs with a LOS perpendicular to the accretion disk, which might not be the case. Indeed, the UV and X-ray variability detected for many of these LINERs \citep{2005ApJ...625..699M,2015A&A...579A..90H} is in favour of a direct view of the AGN (i.e., perpendicular to the disk under the UM). This is consistent with the fact that most of the [\ion{O}{iii}] morphologies found for LINERs are spheroids, if we asume the [\ion{O}{iii}] traces the NLR. In addition, the fact that we detected a clear correspondence between soft X-ray and [\ion{O}{iii}] morphologies only in objects with $\log(L_\mathrm{HX})>40$, and also that all the objects where soft X-rays and [\ion{O}{iii}] match their morphologies seem to better follow the previously found relation between the size of the region and the hard X-ray luminosity (see Fig.~\ref{fig:hard_r} and Section~\ref{sec:oiii_morph}), may argue in favour of the scenario in which the AGN does not have enough thrust to ionise in the low luminosity regime \citep{2006ApJ...648L.101E,2009ApJ...701L..91E}, ruling out photoionisation by the AGN at both soft X-ray and [\ion{O}{iii}] emissions. In this case, the most reasonable explanation for the [\ion{O}{iii}] is the host galaxy emission, which anyhow could also be on top of the AGN, preventing its detection and erasing the connection \citep{2014A&A...567A..92G}. The host galaxy can contribute either as star formation or shocks to the total [\ion{O}{iii}] emission. Regarding the soft X-ray origin, \citet{2014MNRAS.440..269M} confirmed that jets are the main responible of soft X-ray emisison for their sources. In our sample, jets are identified in NGC\,1052 \citep{2004A&A...426..481K}, where the jet position angle would be consistent with the extended soft X-ray emission shown here.

\section{Summary and conclusions}
\label{sec:summary}
We compared the soft X-ray and [\ion{O}{iii}] emissions of active galaxies to test whether they match in different optical classes of AGN. Our sample contains nine type-1 Seyferts, 10 type-2 Seyferts, and eight LINERs. In summary, for the sample studied here, we found: 

\begin{itemize}
\item The soft X-ray and [\ion{O}{iii}] morphologies are both extended in AGN.
\item The soft X-ray morphologies are mainly spheroidal in type-1 Seyferts (67\%), conical in type-2 Seyferts (80\%), and both spheroidal and conical in LINERs.
\item The [\ion{O}{iii}] morphologies in type-1 Seyferts are both spheroidal and conical, conical in type-2 Seyferts (90\%), and spheroidal in LINERs (88\%). The percentage of spheroidal morphologies is higher in type-1 than in type-2 Seyferts.
\item A good match between the soft X-ray and [\ion{O}{iii}] morphologies was found in all the type-2 Seyferts. However, it is less frequent (22\%) for type-1 Seyferts and it is not seen at all in LINERs. This match between morphologies is independent on the continuum subtraction from the [\ion{O}{iii}] image indicating that the [\ion{O}{iii}] emission line dominates the emission within the filter.
\item The hard X-ray emission is mainly point-like in our sample. However, we found four objects (type-2 Seyferts) showing evidence of extended emission in this regime too.
\item The contribution of the [\ion{O}{iii}] extended luminosity in type-2 Seyferts and LINERs is mainly above 50\% over the total [\ion{O}{iii}] luminosity.
\end{itemize}

The type-2 Seyferts analysed here show a good resemblance between the soft X-ray and [\ion{O}{iii}] emissions. This is an observational evidence for a common physical origin for both of them. In the case of type 1 Seyferts, we discussed that orientation effects could diminish the AGN photoionisation component. Based on the morphologies of the eight LINERs studied here, we discarded a common origin for the [\ion{O}{iii}] and soft X-ray emissions in these objects. Furthermore, high column density and hard X-ray luminosity are important ingredients, linked with sources displaying a good resemblance between the soft X-ray and [\ion{O}{iii}] emissions.

\section*{Acknowledgements}

C.G.G acknowledges support from the Plan Nacional de Investigaci\'on y Desarrollo of the Spanish Ministry of Economy and Competitiveness projects AYA2013-46724-P, AYA2012-30717, and AYA2009-10368, and funding from 2014 Summer Grants Programme at the IAC. C.R.A acknowledges the Ram\'on y Cajal Programme of the Spanish Ministry of Economy and Competitiveness through project RYC-2014-15779. We thank J. Masegosa for kindly providing the fitting parameters of her LINERs study.




\bibliographystyle{mnras}
\bibliography{softoiii_agn.bib}



\appendix

\section{Notes on individual objects}
\label{sec:ind_results}
In this appendix, we present relevant assessments detected in some of the objects in our sample individually. Note that this section is not an attempt to fully describe each object but to give relevant notes for some of the sources in our sample. 

\begin{itemize}

\item[$\rm{\bullet}$]{\underline{Circinus}: We see that the [\ion{O}{iii}] emission is located in the nuclear region and also in several star-forming regions. This emission follows a morphological pattern very different from the continuum structure that traces the general shape of the galaxy. The continuum emission does not match the soft X-ray morphology, but it does the [\ion{O}{iii}] emission in the regions with a high S/N ratio. The soft X-ray emission is widely more extended than the [\ion{O}{iii}] one and we notice that reaches regions away from the continuum emission. The hard X-ray pattern seems to be extended, but our PSF study reveals that it is due to several compact sources.}

\item[$\rm{\bullet}$]{\underline{M31}: The majority of the optical emission comes from the continuum and the [\ion{O}{iii}] contribution is low. In the X-rays we detect two compact sources in the nucleus in the hard X-ray energies that do not match the position of the soft emission.}

\item[$\rm{\bullet}$]{\underline{M51a}: The continuum image shows a clear spiral pattern with star-forming regions in the [\ion{O}{iii}]. The nuclear region also emits in the [\ion{O}{iii}], showing a good match with the soft X-ray emission in the parts with a high S/N level. The soft X-ray morphology is widely more extended, with structures that resemble axial precession in the low brightness areas. In addition, we see two main emission regions forming a cone-shaped area that match the emission in the hard X-ray regime.}

\item[$\rm{\bullet}$]{\underline{M77}: The continuum emission traces the spiral shape of the galaxy, but even with this continuum emission we can start to suspect a resemblance with the soft X-ray morphology. The good match is clearer in the [\ion{O}{iii}]+continuum image and in the continuum-subtracted [\ion{O}{iii}] image even clearer. Nevertheless, as we subtract continuum signal we lose the resemblance in the low brightness parts of the galaxy. The morphological good match is also present between the soft and hard X-ray images in the high S/N regions.}

\item[$\rm{\bullet}$]{\underline{M90}: In this case, the soft X-ray image show extended features with two regions in the nucleus with important emission, one of them presenting hard X-rays too. The other region show also [\ion{O}{iii}] emission. What we see would be a good match between soft X-ray and [\ion{O}{iii}] emissions, but just for a little area of the total extension in the soft X-rays.}

\item[$\rm{\bullet}$]{\underline{M105}: In the X-ray images, we find the same point-like structures both reproduced in the soft and hard X-ray regimes.}

\item[$\rm{\bullet}$]{\underline{NGC\,1365}: This is an example of a soft X-ray emission with a diffuse morphology that does not match neither cone-shaped nor spheroidal categories mentioned in Sect.~\ref{sec:xray_morph}. The emission better resembles the inner regions of the spiral pattern that shapes the galaxy morphology in the continuum.}

\item[$\rm{\bullet}$]{\underline{NGC\,1386}: We present in Fig.~\ref{fig:comp_ngc1386} the complete set of images for this object. We see that the continuum emission is completely different from the [\ion{O}{iii}] and the morphology changes drastically, resembling that of the soft X-rays in the [\ion{O}{iii}] case. The hard X-ray emission is also diffuse and matches the soft one. We can see reproduced three main emission regions in both X-ray images.}

\item[$\rm{\bullet}$]{\underline{NGC\,3393}: This is another excellent example of radical change from continuum to [\ion{O}{iii}] appearance. The [\ion{O}{iii}] morphology is elongated and spiral-shaped, resembling the cone-shaped soft X-ray image. However, this good match between both morphologies is not clear comparing [\ion{O}{iii}]+continuum with the soft X-ray band, making this source a fantastic case to illustrate the importance of continuum subtraction.}

\item[$\rm{\bullet}$]{\underline{NGC\,4051}: The [\ion{O}{iii}] image clearly shows a cone-shaped emission that could be linked to the soft X-ray one. However, the soft X-ray emission is spheroidal as well as in the hard X-ray regime. Similar cases are NGC\,4395 and NGC\,5273.}

\item[$\rm{\bullet}$]{\underline{NGC\,4151}: A typical example of drastically change from continuum to [\ion{O}{iii}] emission resembling the soft X-ray morphology. As shown in Fig.~\ref{fig:subs_ngc4151_ngc4388} the difference from [\ion{O}{iii}]+continuum to pure [\ion{O}{iii}] is minimum.}

\item[$\rm{\bullet}$]{\underline{NGC\,4388}: The continuum emission shows a dust band in the galaxy plane with a cone-shaped [\ion{O}{iii}] morphology. If we compare [\ion{O}{iii}]+continuum with the soft X-ray image, the good match is not clear as the dust band is still present. When we subtract the continuum to produce a pure [\ion{O}{iii}] emission line image, the resemblance with the soft X-ray morphology is complete, as shown in Fig.~\ref{fig:subs_ngc4151_ngc4388}.}

\end{itemize}

\section{Comparison images}
\label{sec:appendix}
In this appendix we compare soft X-ray and continuum-subtracted [\ion{O}{iii}]$\lambda$5007 images for all the objects in our sample. \emph{Chandra} soft X-ray contours are superimposed on \emph{HST} [\ion{O}{iii}] images. The contours correspond to five logarithmic intervals with the minimum value set to 3$\sigma$ over the background and the maximum set to 50\% of the peak value. The optical images are scaled in the same way. The images are centred at the soft X-ray peak. Note that in the case of M31 the scale shown in the upper right corner is set to 10\,pc instead of the 100\,pc for the rest of the objects. Figs.~\ref{fig:app_sy1}, \ref{fig:app_sy2}, and \ref{fig:app_liner} show the sources optically classified as type-1 Seyferts, type-2 Seyferts and LINERs, respectively.

\setcounter{figure}{0}
\begin{figure*}
\includegraphics[width=0.66\columnwidth]{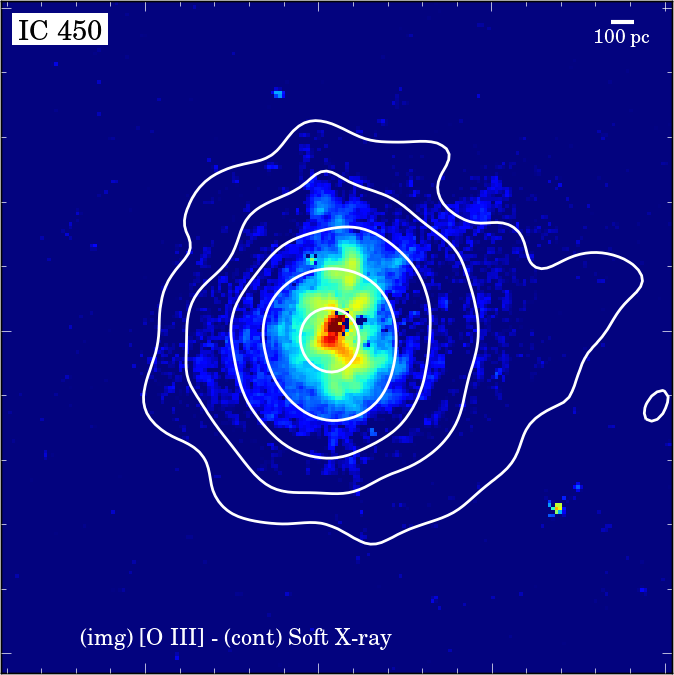}
\vspace{0.1cm}
\hspace{0.1cm}
\includegraphics[width=0.66\columnwidth]{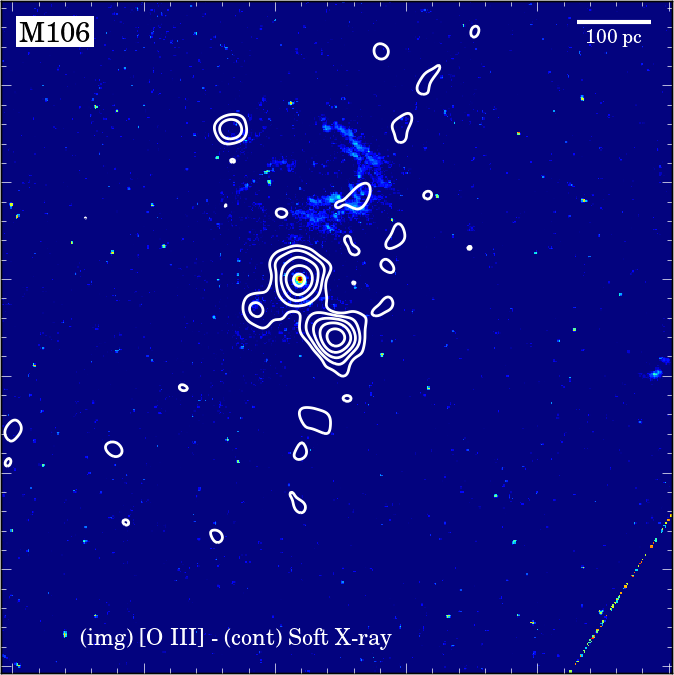}
\vspace{0.1cm}
\hspace{0.1cm}
\includegraphics[width=0.66\columnwidth]{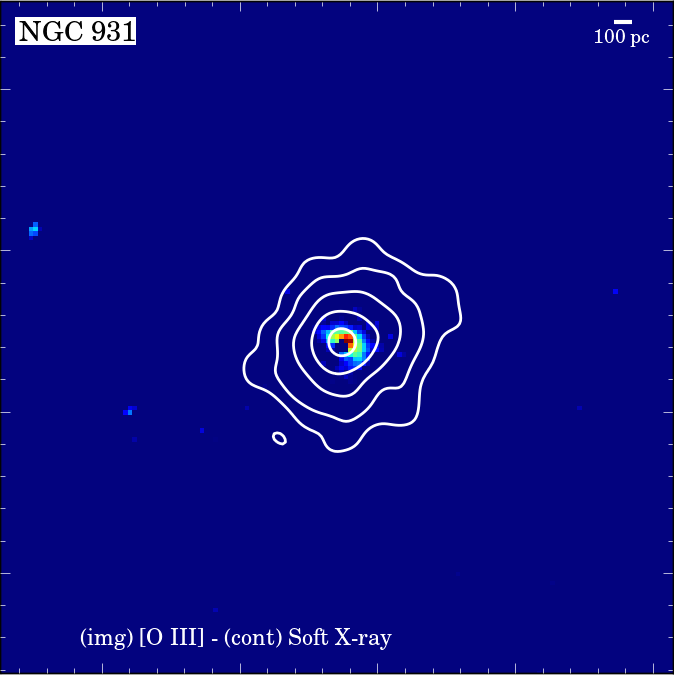}
\vspace{0.1cm}
\includegraphics[width=0.66\columnwidth]{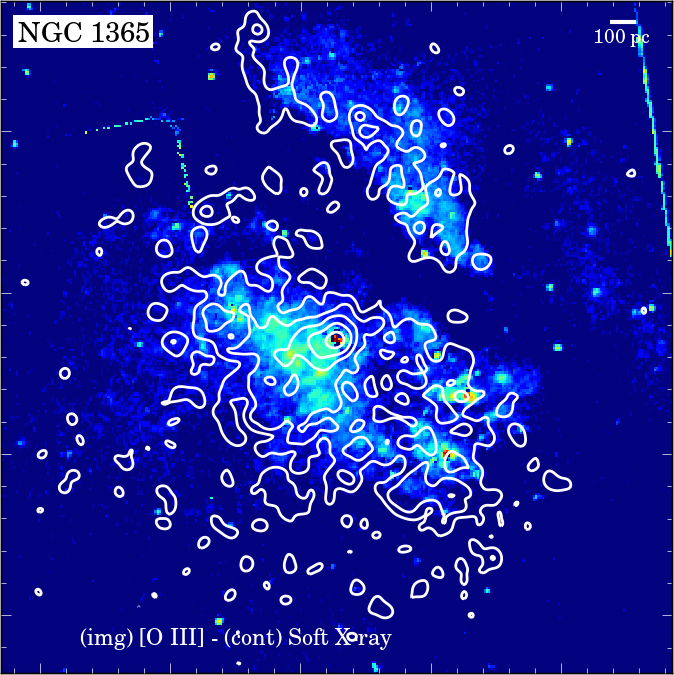}
\vspace{0.1cm}
\hspace{0.1cm}
\includegraphics[width=0.66\columnwidth]{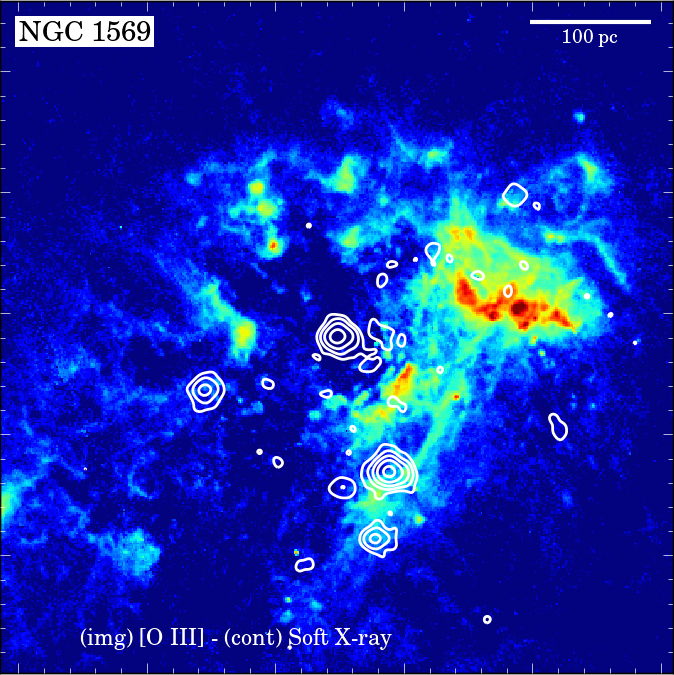}
\vspace{0.1cm}
\hspace{0.1cm}
\includegraphics[width=0.66\columnwidth]{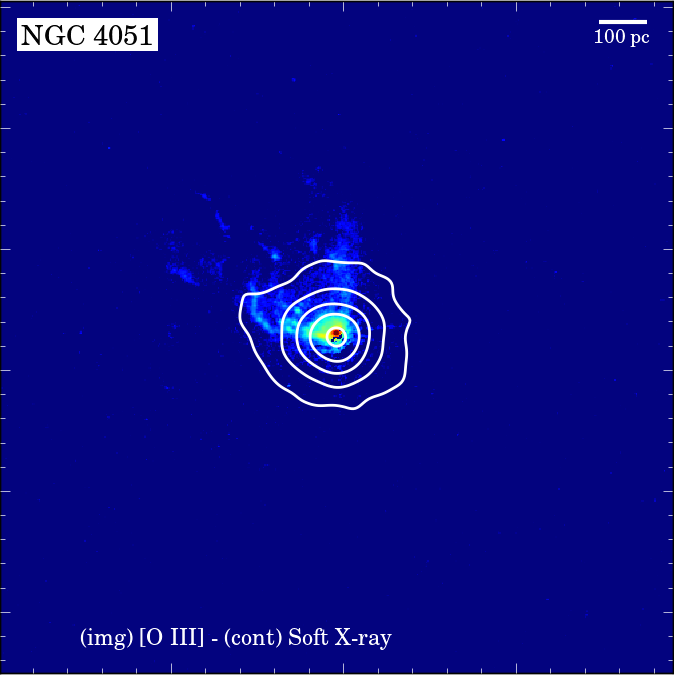}
\vspace{0.1cm}
\includegraphics[width=0.66\columnwidth]{ngc4151_oiii_soft.png}
\vspace{0.1cm}
\hspace{0.1cm}
\includegraphics[width=0.66\columnwidth]{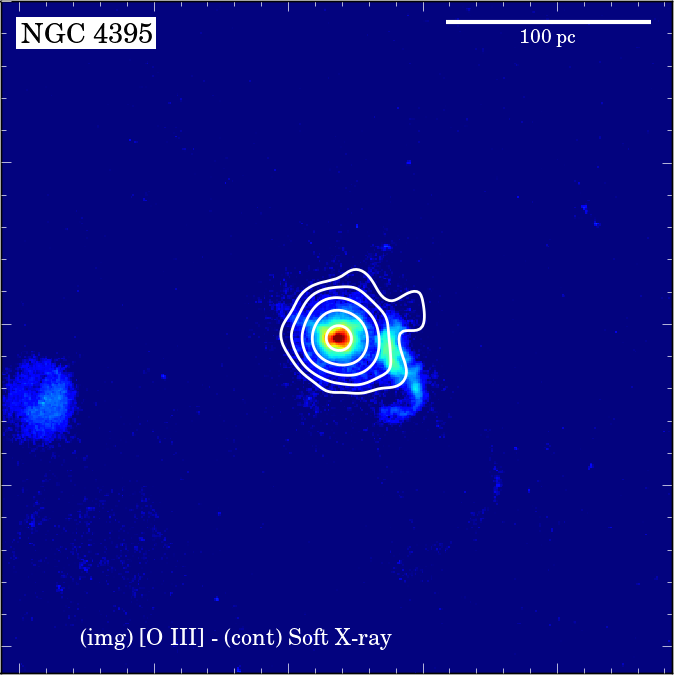}
\vspace{0.1cm}
\hspace{0.1cm}
\includegraphics[width=0.66\columnwidth]{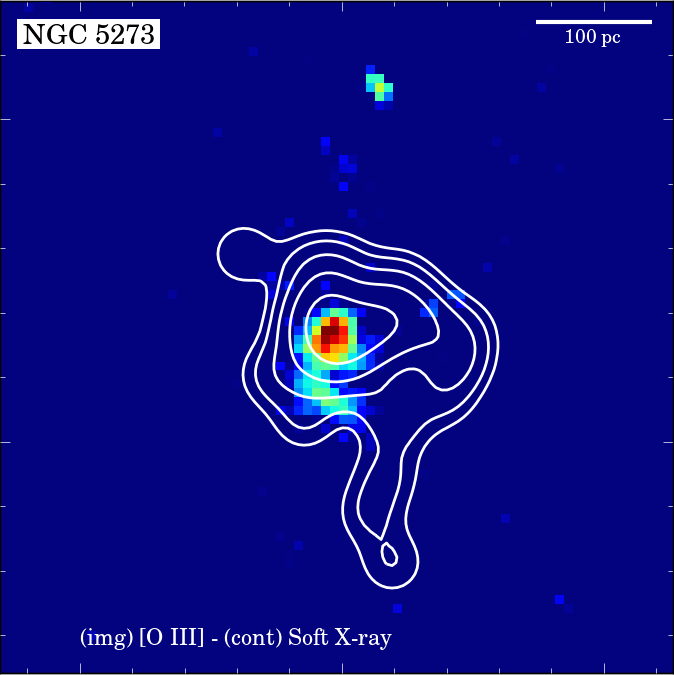}
\vspace{0.1cm}
\caption{\emph{Chandra} soft X-ray contours superimposed on \emph{HST} [\ion{O}{iii}] images for the AGN optically classified as type-1 Seyferts. North is up, East is to the left.}
\label{fig:app_sy1}
\end{figure*}

\begin{figure*}
\includegraphics[width=0.66\columnwidth]{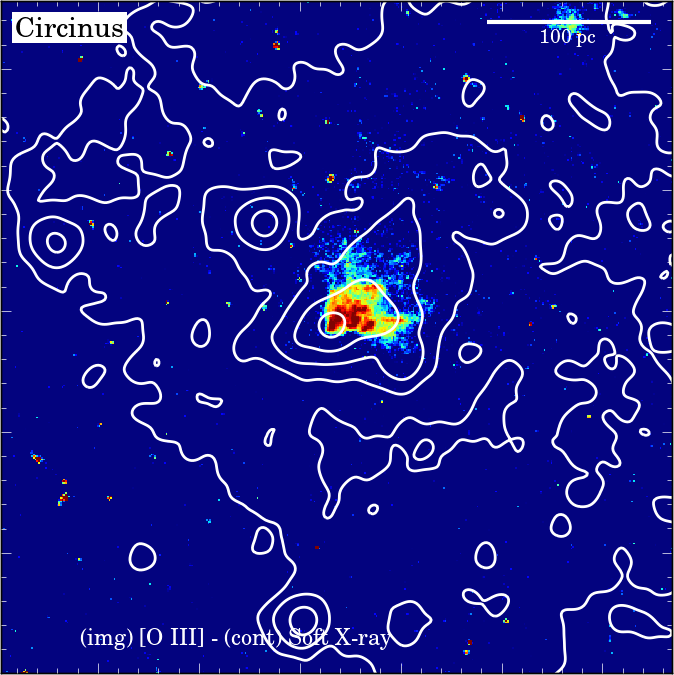}
\vspace{0.1cm}
\hspace{0.1cm}
\includegraphics[width=0.66\columnwidth]{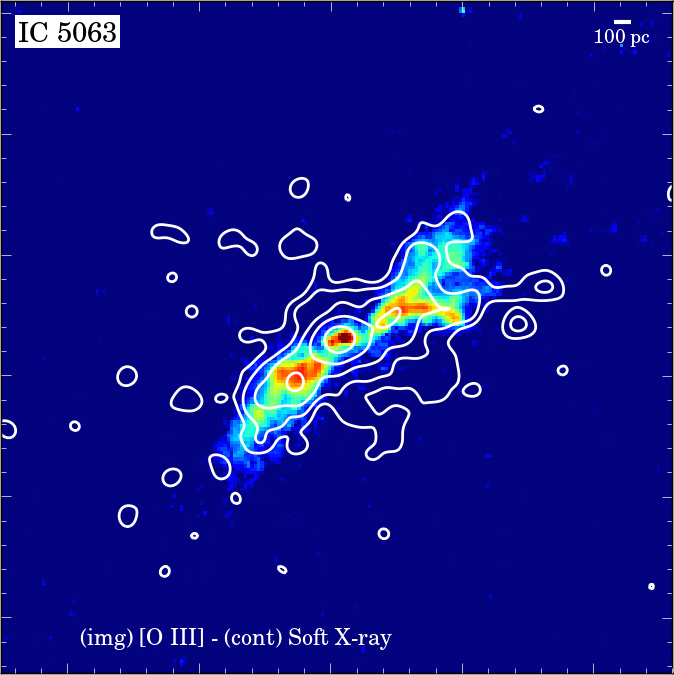}
\vspace{0.1cm}
\hspace{0.1cm}
\includegraphics[width=0.66\columnwidth]{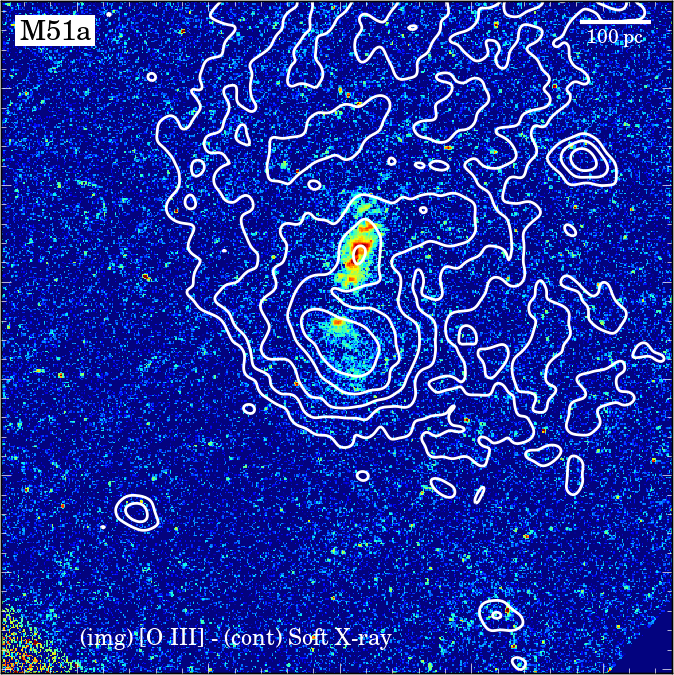}
\vspace{0.1cm}
\includegraphics[width=0.66\columnwidth]{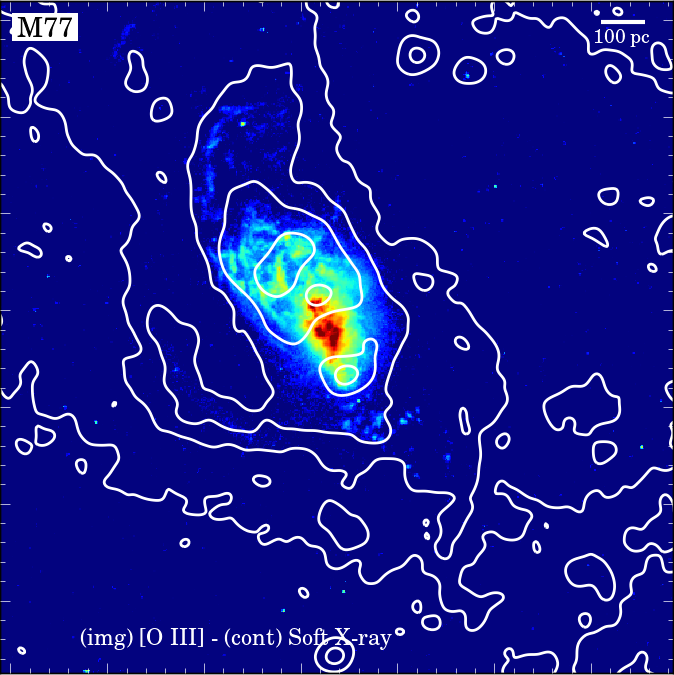}
\vspace{0.1cm}
\hspace{0.1cm}
\includegraphics[width=0.66\columnwidth]{ngc1386_oiii_soft.png}
\vspace{0.1cm}
\hspace{0.1cm}
\includegraphics[width=0.66\columnwidth]{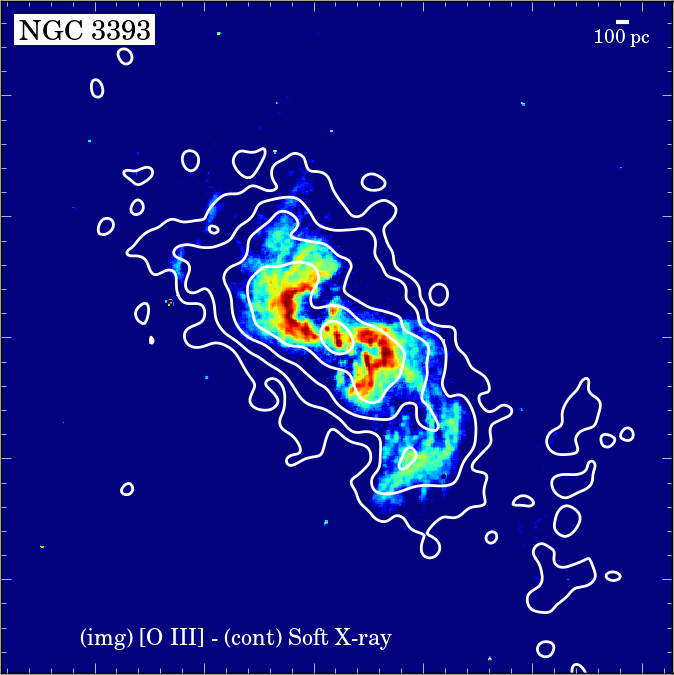}
\vspace{0.1cm}
\includegraphics[width=0.66\columnwidth]{ngc4388_oiii_soft.png}
\vspace{0.1cm}
\hspace{0.1cm}
\includegraphics[width=0.66\columnwidth]{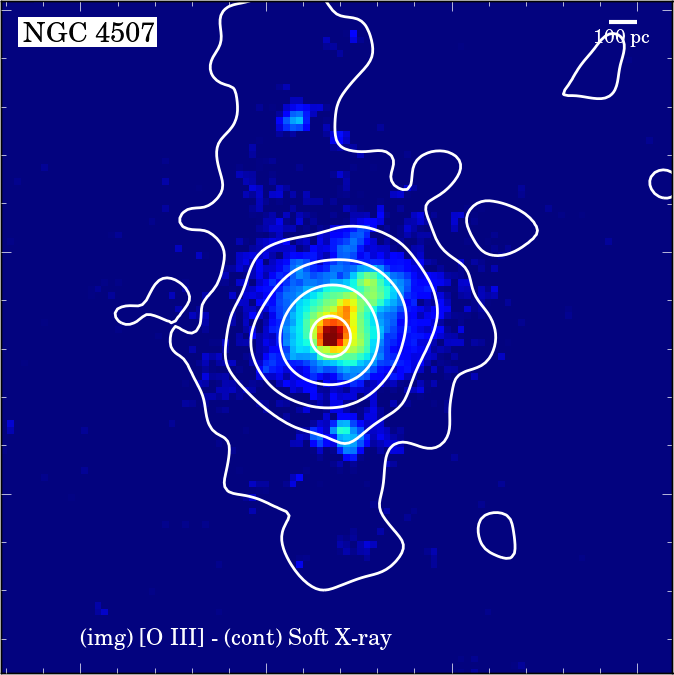}
\vspace{0.1cm}
\hspace{0.1cm}
\includegraphics[width=0.66\columnwidth]{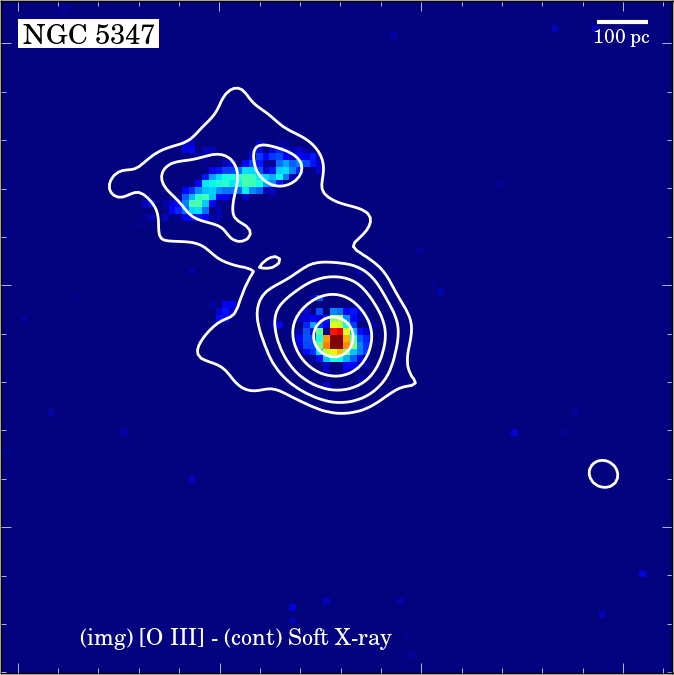}
\vspace{0.1cm}
\includegraphics[width=0.66\columnwidth]{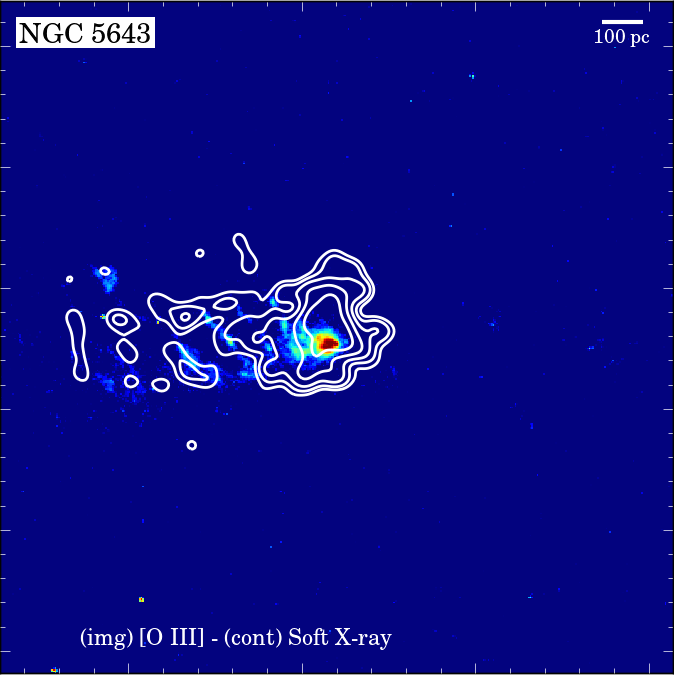}
\vspace{0.1cm}
\caption{\emph{Chandra} soft X-ray contours superimposed on \emph{HST} [\ion{O}{iii}] images for the AGN optically classified as type-2 Seyferts. North is up, East is to the left.}
\label{fig:app_sy2}
\end{figure*}

\begin{figure*}
\includegraphics[width=0.66\columnwidth]{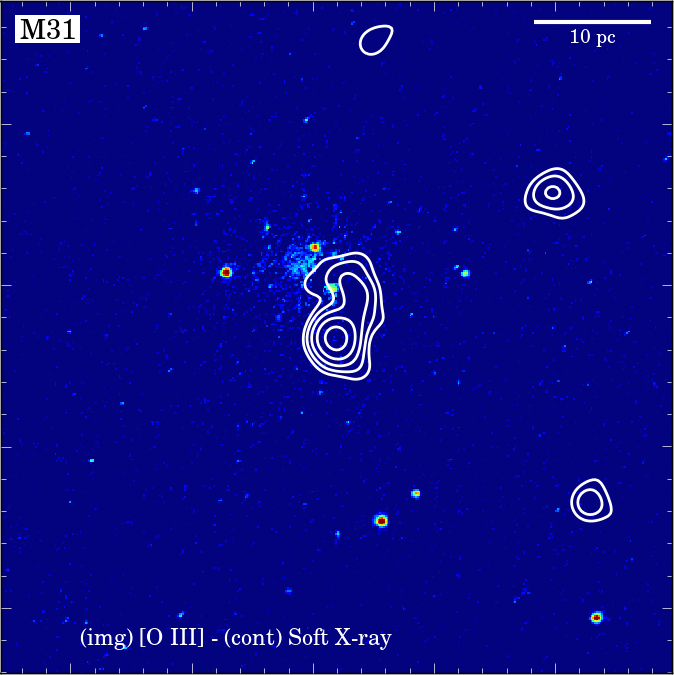}
\vspace{0.1cm}
\hspace{0.1cm}
\includegraphics[width=0.66\columnwidth]{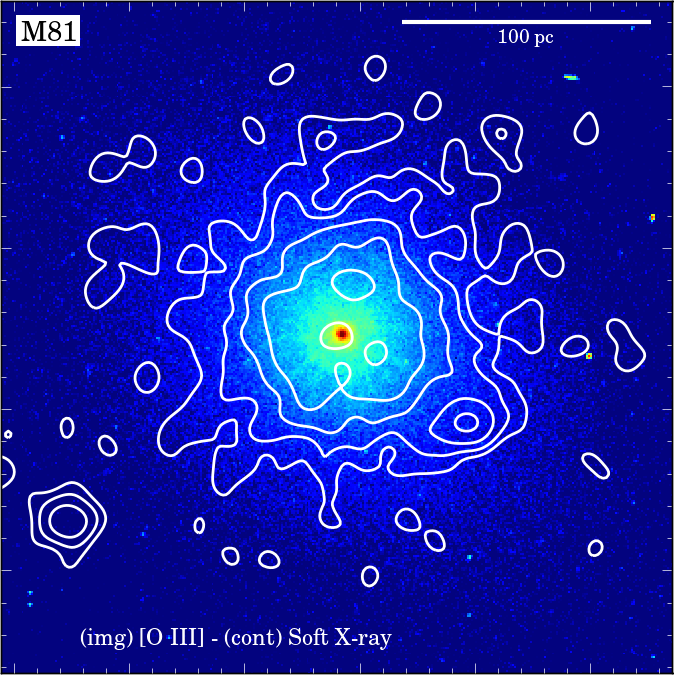}
\vspace{0.1cm}
\hspace{0.1cm}
\includegraphics[width=0.66\columnwidth]{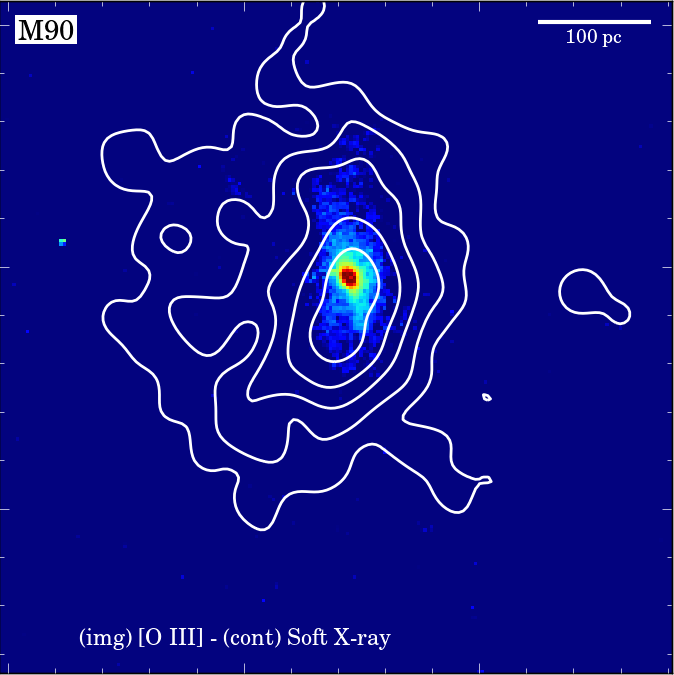}
\vspace{0.1cm}
\includegraphics[width=0.66\columnwidth]{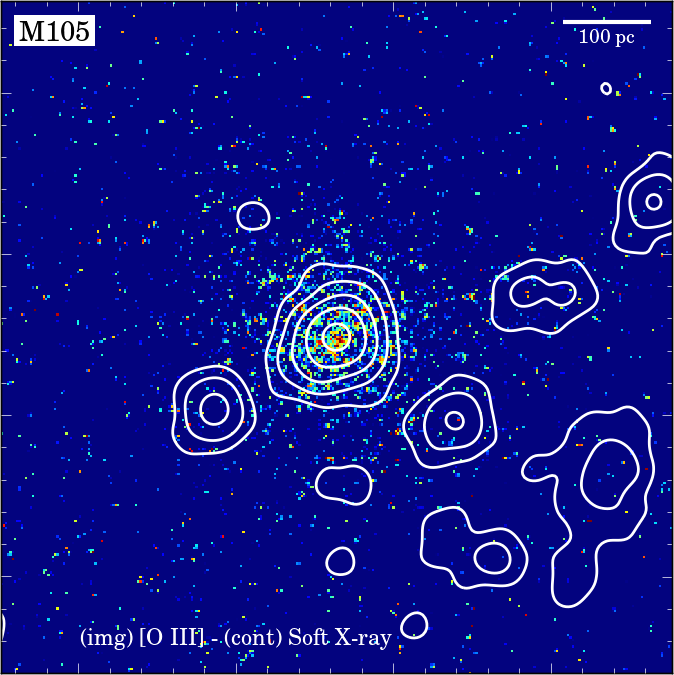}
\vspace{0.1cm}
\hspace{0.1cm}
\includegraphics[width=0.66\columnwidth]{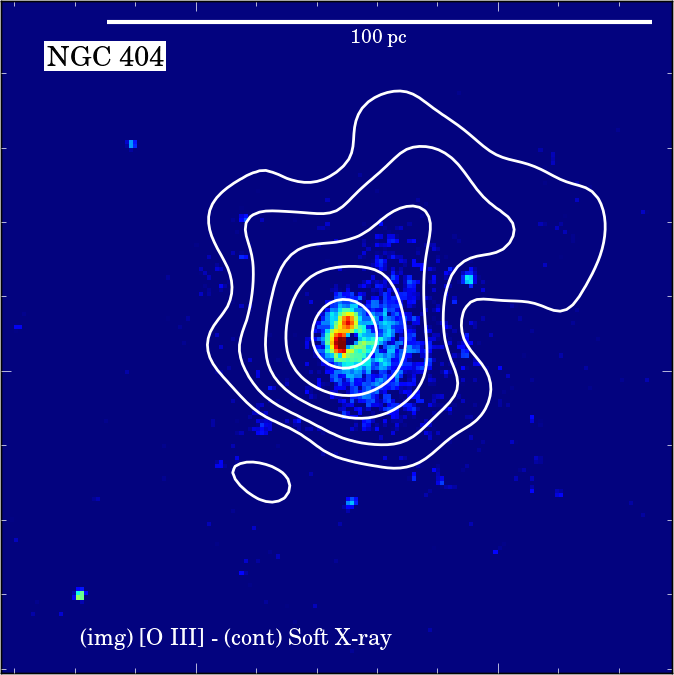}
\vspace{0.1cm}
\hspace{0.1cm}
\includegraphics[width=0.66\columnwidth]{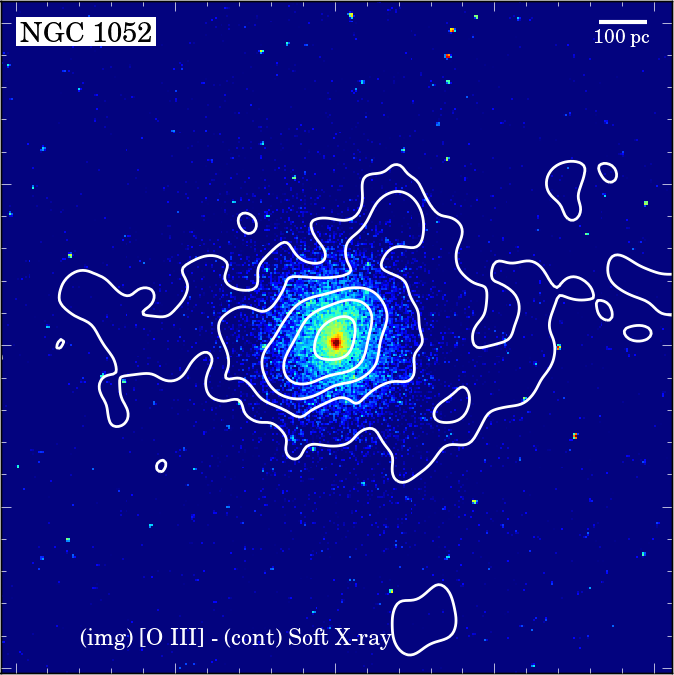}
\vspace{0.1cm}
\includegraphics[width=0.66\columnwidth]{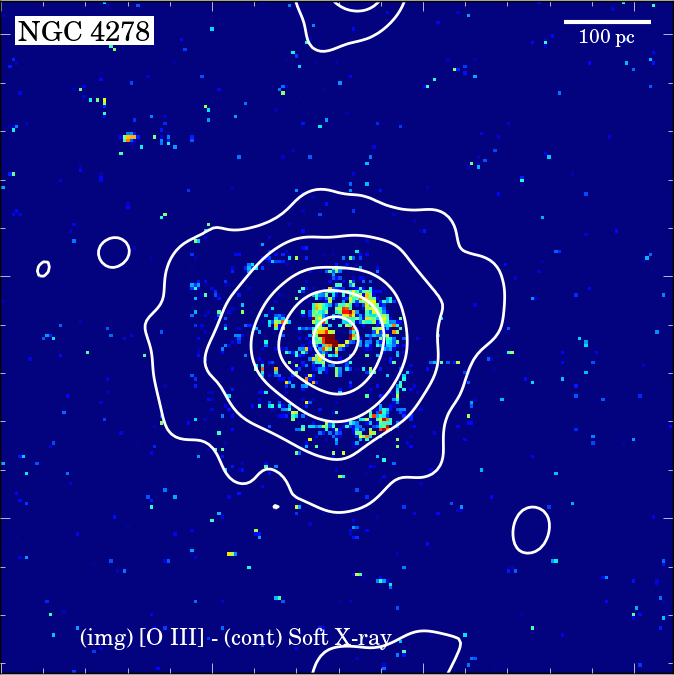}
\vspace{0.1cm}
\hspace{0.1cm}
\includegraphics[width=0.66\columnwidth]{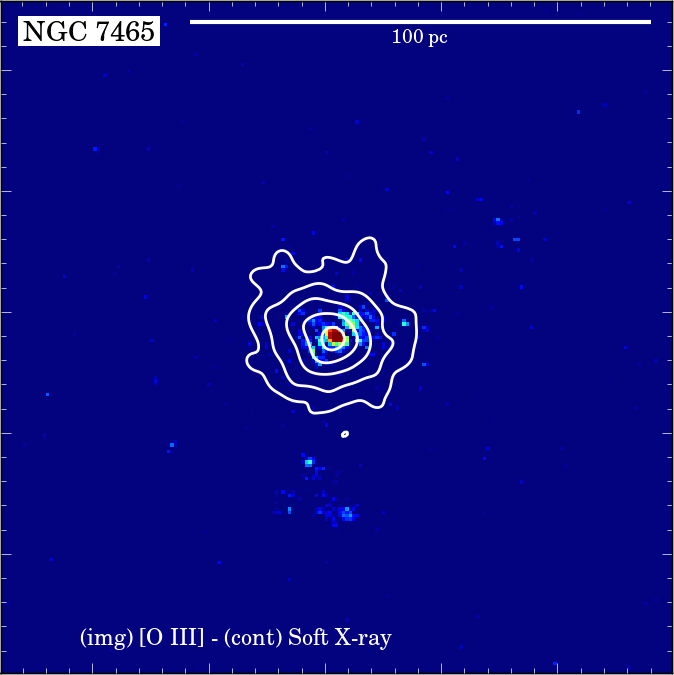}
\vspace{0.1cm}
\caption{\emph{Chandra} soft X-ray contours superimposed on \emph{HST} [\ion{O}{iii}] images for the AGN optically classified as LINERs. North is up, East is to the left.}
\label{fig:app_liner}
\end{figure*}


\bsp	
\label{lastpage}
\end{document}